%% file: FairPlannerTVCG.tex
\documentclass[10pt,journal,compsoc]{IEEEtran}

\ifCLASSOPTIONcompsoc
  \usepackage[nocompress]{cite}
\else
  \usepackage{cite}
\fi
\usepackage{graphicx}      
\usepackage{amsmath,bm,amssymb} 
\usepackage{color}
\usepackage{subcaption}
\usepackage{algorithm}
\usepackage{algpseudocode}%
\usepackage{enumitem}

\DeclareMathOperator*{\argmin}{{\arg\!\min}}
\usepackage{cuted}
\usepackage{capt-of}
\usepackage{multirow}
\usepackage{array}
\newcolumntype{L}{>{\centering\arraybackslash}m{0.8cm}}

\begin{document}
%
\title{IF-City: Intelligible Fair City Planning
\\to Measure, Explain and Mitigate Inequality}
%
%
%
%

\author{
Yan~Lyu,~
Hangxin~Lu,~
Min~Kyung~Lee,
Gerhard~Schmitt,
 and Brian~Y.~Lim
\IEEEcompsocitemizethanks{
\IEEEcompsocthanksitem Yan Lyu is with the School of Computer Science and Engineering, Southeast University, Nanjing, China. E-mail:lvyanly@gmail.com.
\IEEEcompsocthanksitem Hangxin Lu and Gerhard Schmitt are with Future Cities Laboratory, Singapore-ETH Centre, Singapore, and Department of Architecture, ETH Zurich, Switzerland. E-mail: luhangxin@gmail.com, schmitt@arch.ethz.ch.
\IEEEcompsocthanksitem Min Kyung Lee is with the School of Information, The University of Texas at Austin, US. E-mail: minkyung.lee@austin.utexas.edu

\IEEEcompsocthanksitem Brian Y. Lim (Corresponding) is with the School of Computing, National University of Singapore, Singapore. E-mail: brianlim@comp.nus.edu.sg.

}
}

%
%

\markboth{IEEE Transactions on Visualization and Computer Graphics}%
{}


\IEEEtitleabstractindextext{%


\begin{abstract}

With the increasing pervasiveness of Artificial Intelligence (AI), 
many visual analytics tools have been proposed to examine fairness, but they mostly focus on data scientist users.
Instead, tackling fairness must be inclusive and involve domain experts with specialized tools and workflows.
Thus, domain-specific visualizations are needed for algorithmic fairness.
Furthermore, while much work on AI fairness has focused on predictive decisions, less has been done for fair allocation and planning, which require human expertise and iterative design to integrate myriad constraints.
We propose the Intelligible Fair Allocation (IF-Alloc) Framework that leverages explanations of causal attribution (Why), contrastive (Why Not) and counterfactual reasoning (What If, How To) to aid domain experts to assess and alleviate unfairness in allocation problems.
We apply the framework to fair urban planning for designing cities that provide equal access to amenities and benefits for diverse resident types. 
Specifically, we propose an interactive visual tool, Intelligible Fair City Planner (IF-City), to help urban planners to perceive inequality across groups, identify and attribute sources of inequality, and mitigate inequality with automatic allocation simulations and constraint-satisfying recommendations. 
%
We demonstrate and evaluate the usage and usefulness of IF-City on a real neighborhood in New York City, US, with practicing urban planners from multiple countries,
and discuss generalizing our findings, application, and framework to other use cases and applications of fair allocation.

\end{abstract}

\begin{IEEEkeywords}
Fairness, Intelligibility, Explainable Artificial Intelligence, Resource Allocation, Urban Planning
\end{IEEEkeywords}}

\maketitle


\IEEEdisplaynontitleabstractindextext

%
\IEEEpeerreviewmaketitle

\input{Introduction}

\input{RelatedWork}

\input{ConceptualPipeline}

\input{Methods}
\input{VisualDesign}

\input{SystemArchitecture}

\input{CaseStudy}

\input{Evaluation}

\input{Discussion}
\input{Conclusion}




\ifCLASSOPTIONcaptionsoff
  \newpage
\fi



\bibliographystyle{IEEEtran}
\bibliography{reference}
%
\begin{IEEEbiography}[{\includegraphics[width=1in,height=1.25in,clip,keepaspectratio]{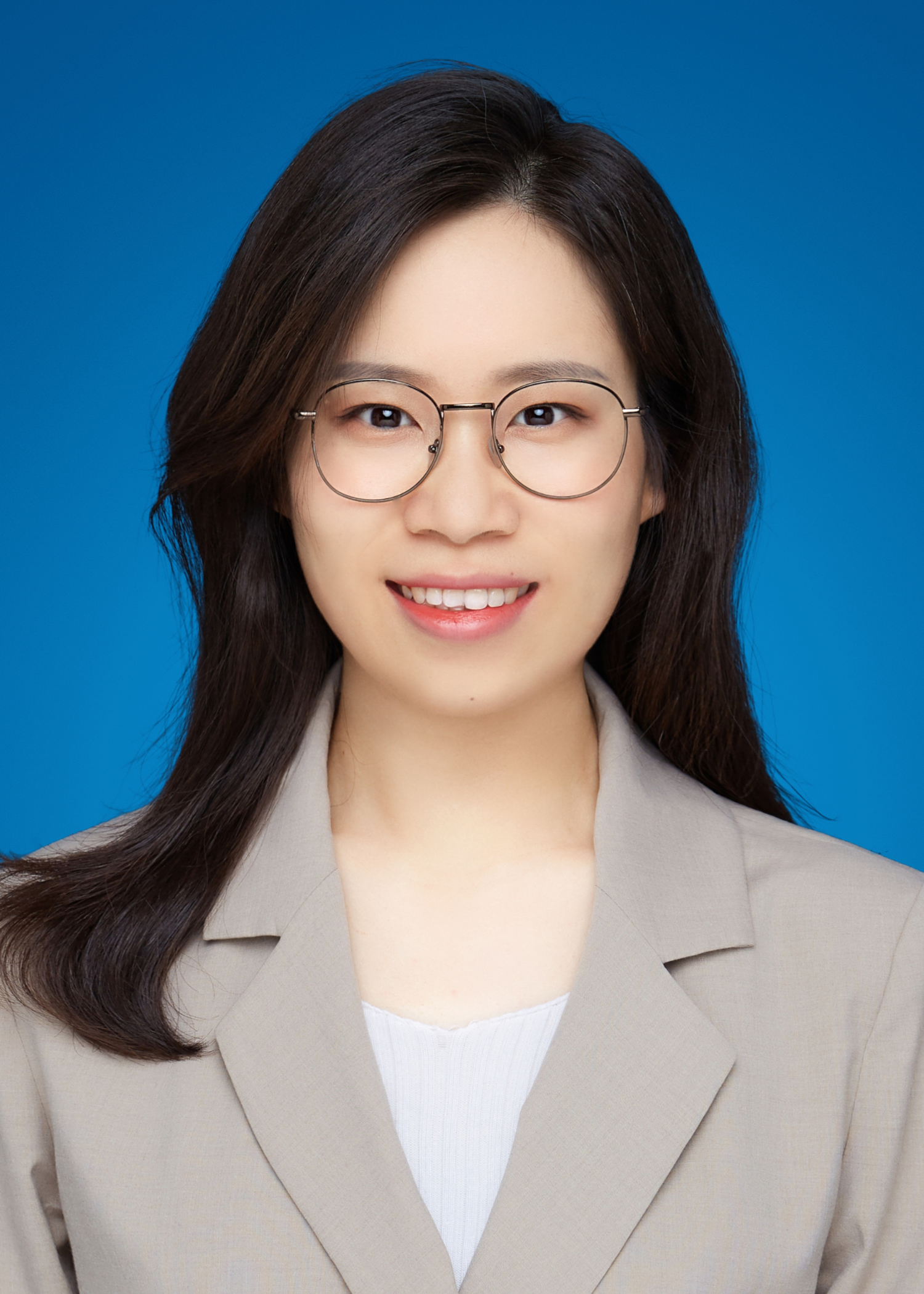}}]{Yan Lyu} received the Ph.D. degree in computer science from City University of Hong Kong, Hong Kong, in 2016, and the M.S. degree in pattern recognition and intelligent systems from University of Science and Technology of China, China, in 2013. She was a Postdoctoral Research Fellow with Hong Kong Baptist University, Hong Kong, in 2017, and with National University of Singapore, Singapore, from 2017 to 2020. She is an Associate Professor in the School of Computer Science and Engineering at Southeast University, China. Her research interests include data analytics and visualization, spatio-temporal data mining, and smart city.
\end{IEEEbiography}
\begin{IEEEbiography}[{\includegraphics[width=1in,height=1.25in,clip,keepaspectratio]{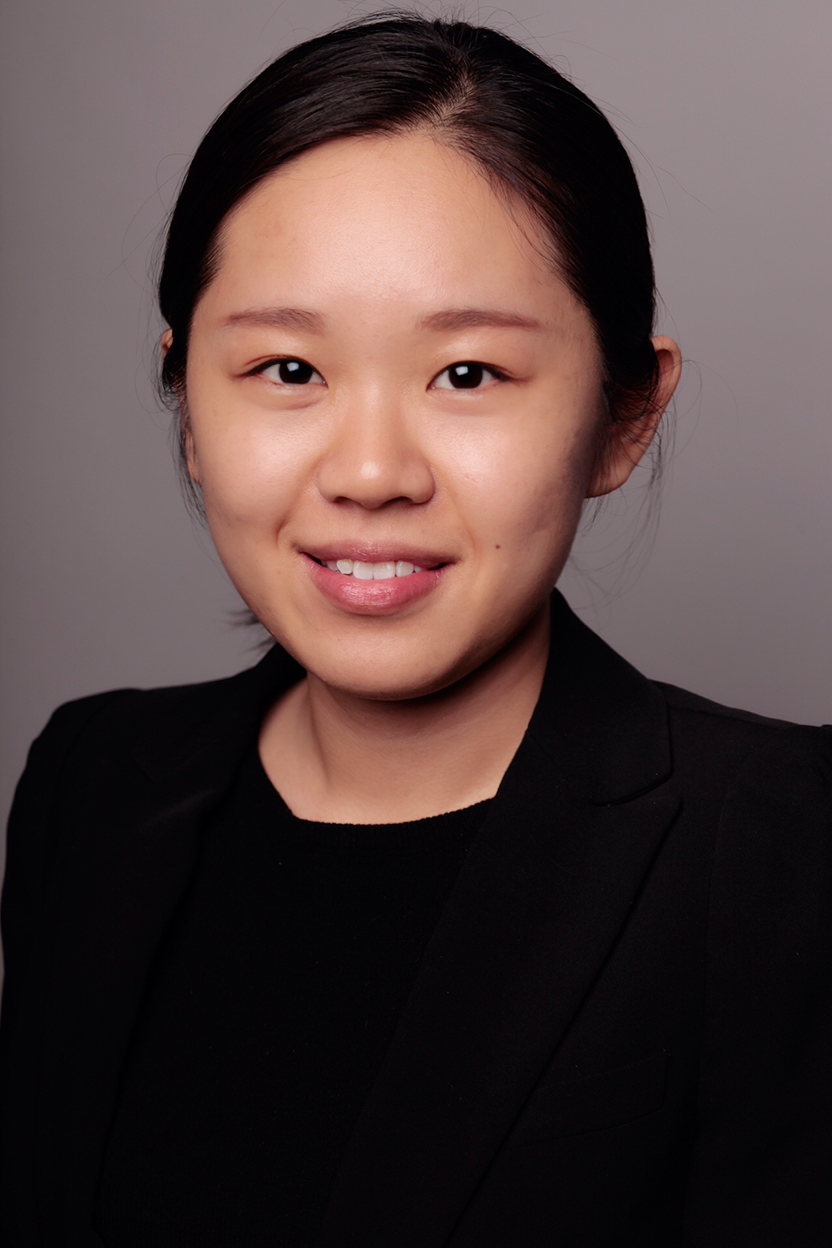}}]{Hangxin Lu} received the M.Sc. degree in Electrical Engineering from EPFL, Lausanne, Switzerland, in 2016, and the PhD degree from ETH Zurich, Switzerland. She is currently a technology consultant working on big data governance. Her research interests include interactive data visualization, participatory design, smart and responsive cities, and digital urban planning applications.
\end{IEEEbiography}
\begin{IEEEbiography}[{\includegraphics[width=1in,height=1.25in,clip,keepaspectratio]{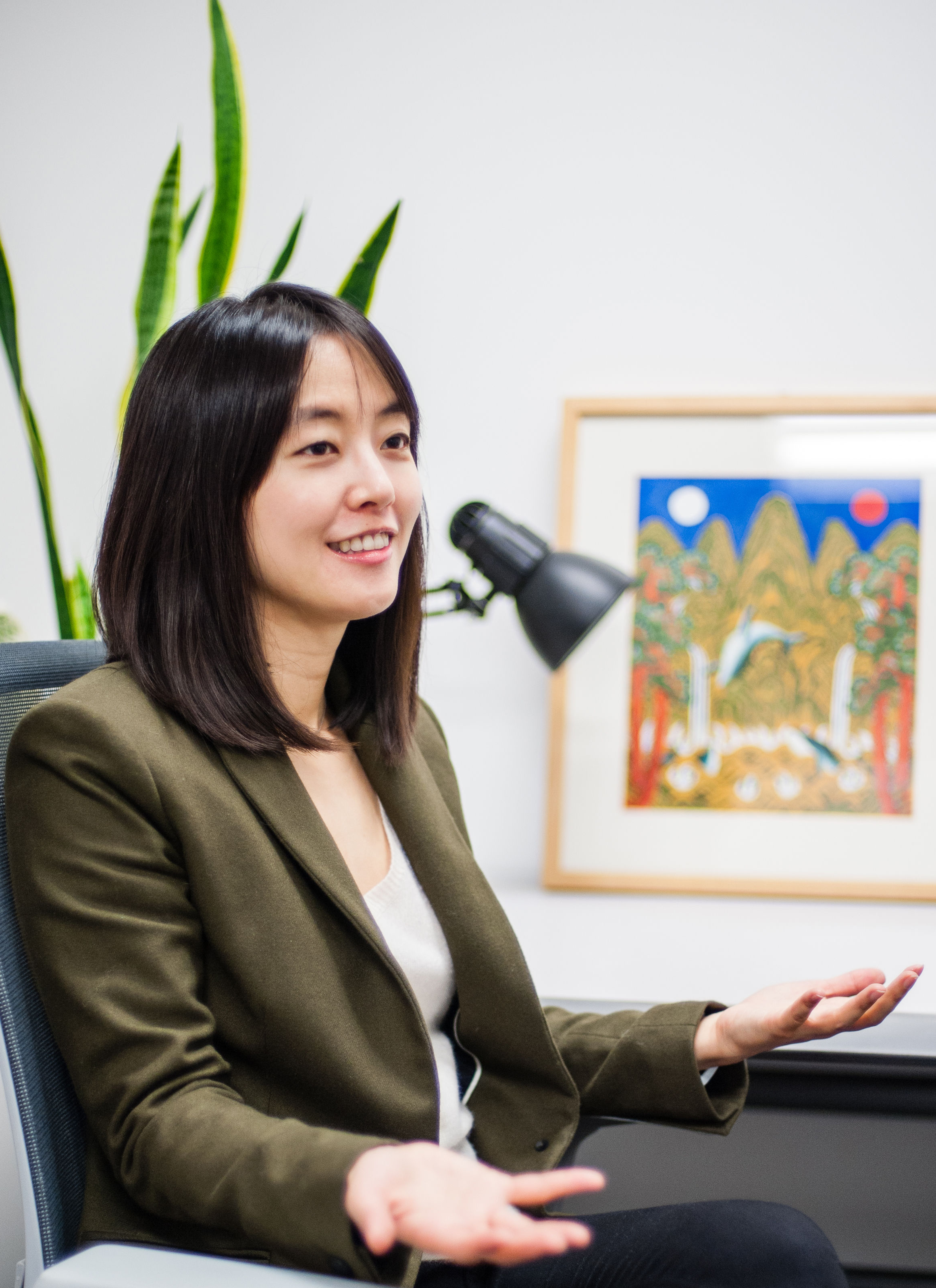}}]{Min Kyung Lee} is an assistant professor in the School of Information at the University of Texas at Austin. She received a Ph.D. in Human-Computer Interaction from Carnegie Mellon University. She has conducted some of the first studies that empirically examine the social implications of algorithms’ emerging roles in management and governance in society. Her current research interests include human-centered AI with a focus on human-centered perspectives on AI fairness and equity, and participatory design methods for community-centered AI design.
\end{IEEEbiography}
\begin{IEEEbiography}[{\includegraphics[width=1in,height=1.25in,clip,keepaspectratio]{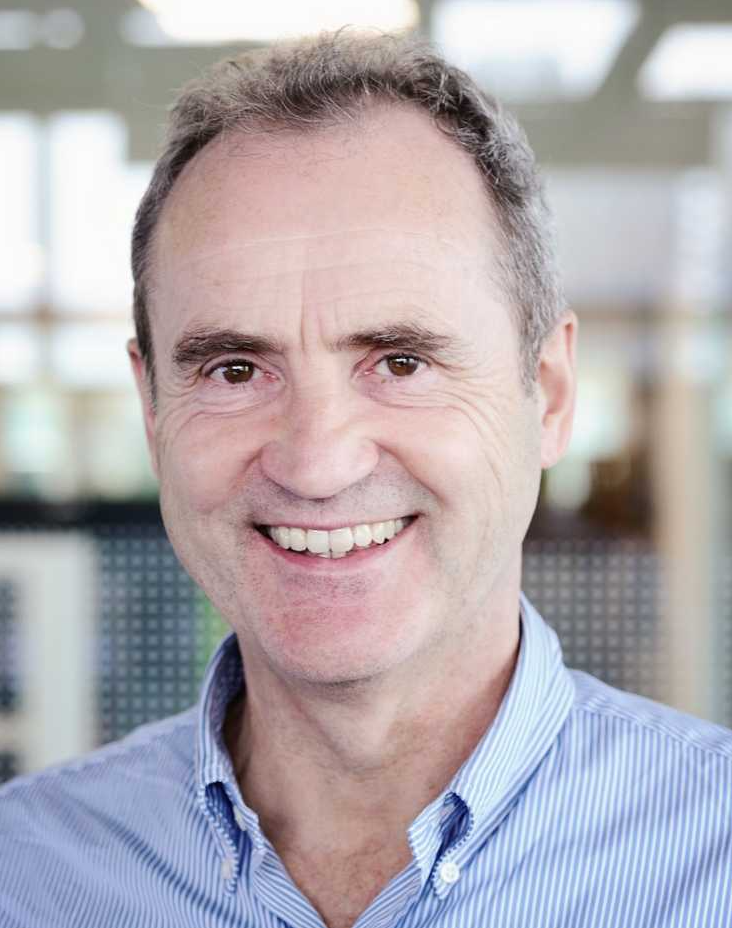}}]{Gerhard Schmitt} is Director of the Singapore-ETH Centre in Singapore, Lead PI of the ETH Future Cities Responsive Cities Scenario, Lead PI of the Cooling Singapore Project and Professor Emeritus of Information Architecture at ETH Zurich. He holds a Dipl.-Ing. and a Dr.-Ing. degree from the Technical University of Munich and a Master of Architecture from the University of California, Berkeley. His research focuses on urban climate design, big data informed urban design, urban metabolism, smart and responsive cities, simulation and visualization.
\end{IEEEbiography}
\begin{IEEEbiography}[{\includegraphics[width=1in,height=1.25in,clip,keepaspectratio]{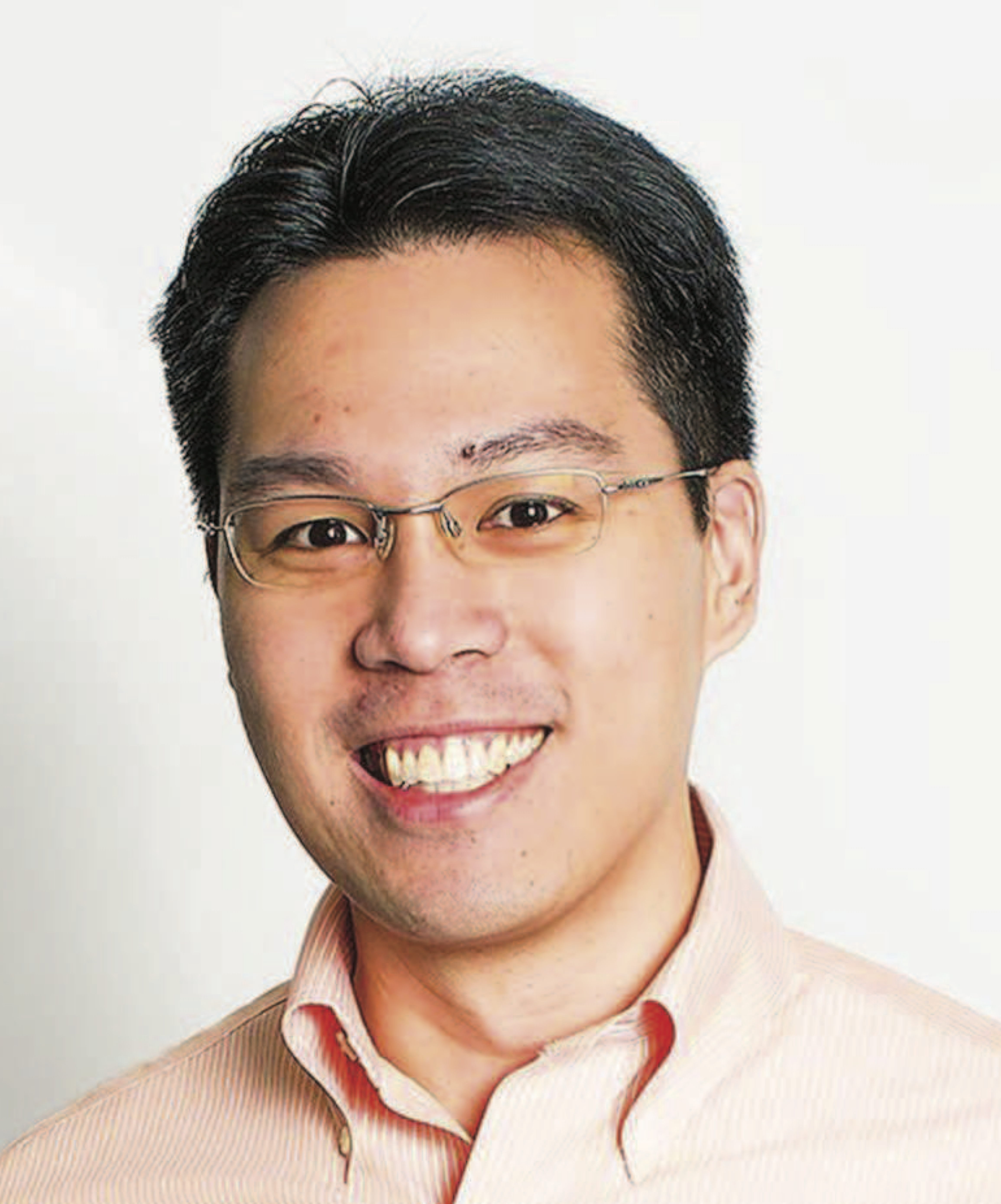}}]{Brian Y. Lim}
received the BS degree in engineering physics from Cornell University, Ithaca, New York, in 2006, and the PhD degree in human-computer interaction from Carnegie Mellon University, Pittsburgh, Pennsylvania, in 2012. He is currently an assistant professor with the Department of Computer Science, National University of Singapore (NUS). His research interests include ubiquitous computing, explainable artificial intelligence, interfaces, and applications for urban data analytics and smart healthcare. 
\end{IEEEbiography}

\vfill




\end{document}

%% file: Introduction.tex
\section{Introduction}

\begin{figure*}
\centering
\includegraphics[width=1\textwidth]{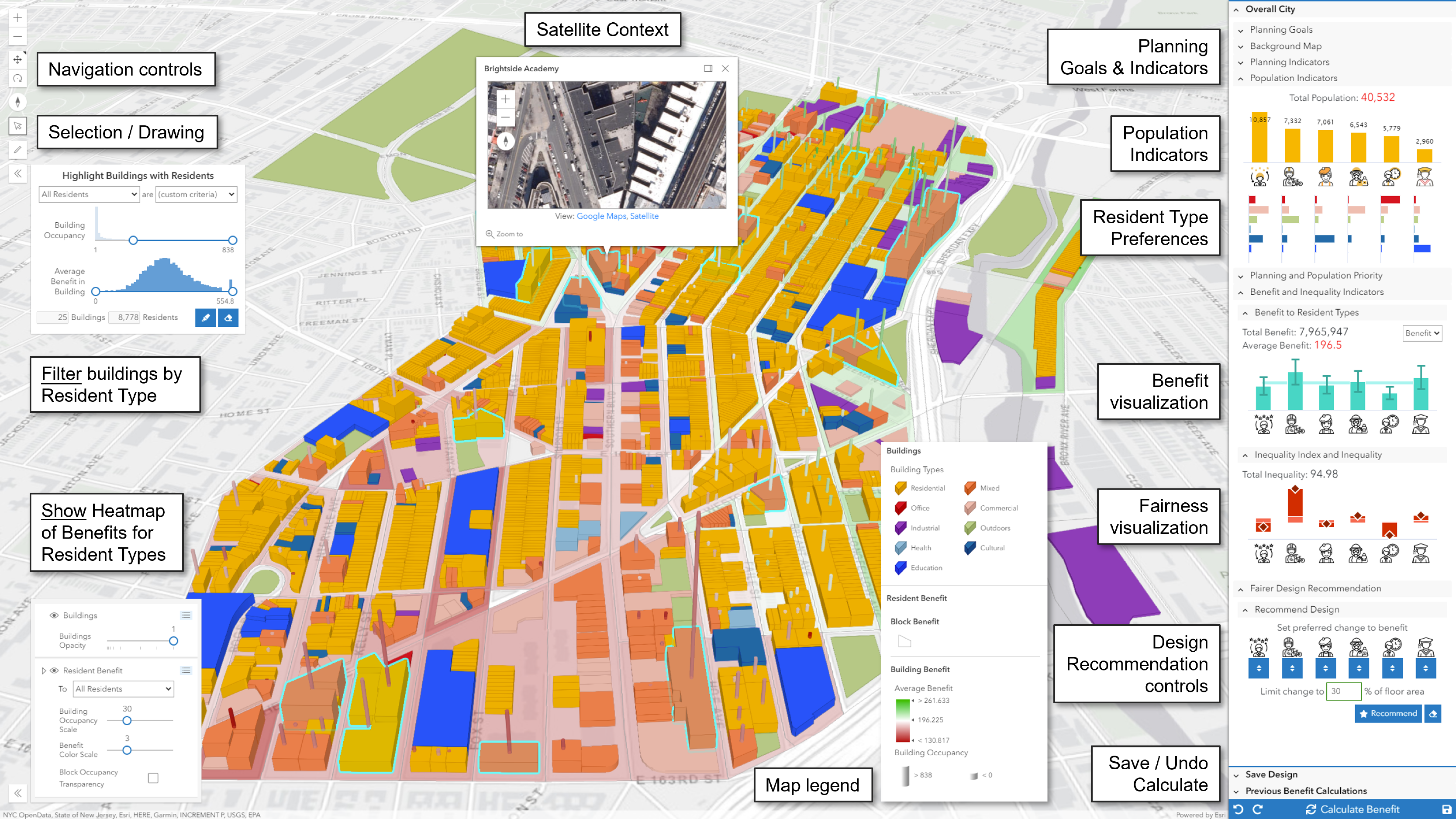}
\vspace{-0.25in}
\captionof{figure}{
Dashboard user interface of IF-City for fair urban planning with main 3D map view, side bar with several indicators, and pop-up controls.
Urban planners can add or edit buildings of different types, and evaluate the impact on resident benefits and inequality.
Users can also query for recommendations to improve fairness and save design iterations.
}
\vspace{-0.08in}
\label{fig:DesignDashboard}
\end{figure*}

The increasing pervasiveness of Artificial Intelligence (AI) has raised concerns of algorithmic bias, unfairness and discrimination as AI benefits some segments of society while disadvantaging others~\cite{buolamwini2018gender}. 
Current methods to assess fairness in model outcomes and decisions focus on visualizing tabular data with bar charts and scatter plots~\cite{ahn2019fairsight,cabrera2019fairvis}, which are well suited for data scientists and machine learning engineers.
However, assessing fairness requires domain experts and stakeholders to consider relevant factors and balance multiple criteria~\cite{wang2020visual}. Indeed, many real-world applications require tailored, domain-specific visualizations, such as time series visualizations for financial\cite{dang2012timeseer}, sensor or health applications\cite{zhang2018idmvis}, image-based visualizations for medical radiology\cite{dmitriev2019visual}, and maps and charts for urban and transportation planning\cite{liu2016smartadp,lyu2019od}. Hence, instead of having just data scientists assess model fairness, fairness visualizations should be integrated into domain-specific tools to involve other stakeholders to participate collaboratively~\cite{lee2019webuildai,robertson2021modeling}.

Such participatory design will require users to understand how the intelligent system determines fairness in order for them to mitigate unfairness.
Fortuitously, explanations are an effective way for users to understand and trust system outcomes~\cite{abdul2018trends, guidotti2018survey, wang2019designing}. Recently, many explainable AI (XAI) techniques have been developed, such as attribution~\cite{lundberg2017unified, ribeiro2016should}, contrastive~\cite{miller2019explanation} and counterfactual~\cite{wachter2017counterfactual,wexler2019if} explanations. 
This motivates us to apply XAI techniques to help planners and designers understand how fairness is computed and how to improve it in their planned solutions.

Much recent progress on fair machine learning focuses on algorithmic bias in inference-based decisions \cite{corbett2018measure,mehrabi2021survey}, yet there is also a strong need for fairness in resource allocation for domains such as urban planning, organizational management, public services, and healthcare~\cite{lee2019webuildai}. 
Automatic algorithms can find optimal allocations, but they may overlook nuances and normative decision points that are part of complex real world systems~\cite{robertson2021modeling}; instead, the participatory involvement of stakeholders and domain experts is necessary in fair allocation decisions~\cite{lee2019webuildai}.

Hence, we propose the Intelligible Fair Allocation (IF-Alloc) Framework that leverages explainable AI to help users assess fairness, identify sources of inequality, and mitigation inequality. 
Specifically, we adopt Lim and Dey's user-centered intelligibility framework of question types \cite{lim2009assessing} to satisfy user reasoning needs for causal attribution (Why), contrastive (Why Not) and counterfactual reasoning (What If, How To)~\cite{miller2019explanation,wang2019designing}. 
We use the Generalized Entropy Index as the fairness metric and leverage its additive decomposibility to attribute inequality to constituent components. 

We applied the framework to develop the Intelligible Fair City Planner (IF-City) for fair urban planning to balance the benefits from building locations across residents with different preferences. For example, having large parks in the city is valuable to nature lovers, but may cause schools or offices to be placed farther away and disadvantage students and office workers, respectively. 
IF-City is an urban planning visualization and 3D model design tool that helps urban planners to design fairer cities by placing buildings on a map, simulating the allocation of the resident population with iterative proportional fitting (IPF)~\cite{muller2010population}, and computing the fairness indicator for the urban plan based on accessibility to and preferences for different amenities. 
IF-City helps users to understand the fairness calculation by visualizing where benefits were unfairly allocated to different resident types across locations (Why). 
Urban planners can update the urban design to explore how inequalities can be reduced (What If), and compare between design iterations (Why Not). 
IF-City also provides recommendations for editing the quantity and distribution of amenities to optimize fairness (How To), predicts their partial contributions, as calculated by Shapley values, towards improving overall fairness (Why of How To). 
In summary, our contributions are:
\begin{itemize}[leftmargin=0.16in,itemsep = 0in,topsep=0.01in]
\item Intelligible Fair Allocation (IF-Alloc) Framework to use intelligibility features for perceiving fairness, explaining sources of allocative inequality, and mitigating inequality.
\item Intelligible Fair City Planner (IF-City), an interactive visualization tool for urban planners to design fair neighborhoods by interactively and iteratively assess and improve fairness for different resident types.
It allocates residents based on the urban design, and calculates and explains accessibility, benefits, and inequality indicators.
\item A constrained counterfactual recommendation technique to generate recommendations for design edits to conveniently improve fairness in urban planning.
\end{itemize}

We demonstrated IF-City with a use case for designing a neighborhood in New York City to improve fairness, and 
evaluated it in a formative study with practicing urban planners and urban designers to investigate how the explanation features are used to understand and improve fairness in urban designs. 
Users could identify sources of inequality and mitigate inequality through trial and error and with the automatic recommendation. We conclude with a discussion to generalize IF-City and the IF-Alloc framework.

%% file: RelatedWork.tex
\section{Related Work}

We introduce notions of fairness for predictive and allocation systems, and relate them to urban planning. We then describe visualization tools to assess and improve fairness, explain data and models, and analyze and plan cities.

\subsection{Fairness definitions and metrics}

\subsubsection{Metrics of fair outcomes}

Recently, research on bias and fairness in machine learning has been very active~\cite{mehrabi2021survey}.
Such works frame fairness by individual or group fairness. Individual fairness requires that \textit{similar} individuals should have similar outcomes (e.g., people with similar credit scores should have similar loan approval chances). However, defining what makes individuals similar may be difficult. Instead, individuals can be grouped based on having similar attributes or demographics. In socially-sensitive applications, groups are defined by \textit{protected} attributes like age and gender. Group fairness requires that \textit{different} groups have similar outcomes (e.g., younger and older workers with similar other attributes should have similar hiring chances). 
Many fairness metrics (e.g., statistical parity~\cite{dwork2012fairness}, equal opportunity~\cite{Hardt2016}, conditional equality~\cite{beutel2019putting}, intersectional bias~\cite{cabrera2019fairvis}) have been proposed to evaluate the balance in decision outcomes. In this work, we focus on supporting fairness across groups, but focus on resource allocation instead of predictive decisions.

\subsubsection{Metrics of fair allocation}
Unlike algorithmic fairness that considers how fair a prediction result is, \textbf{fair resource allocation} seeks to allocate limited resources fairly to individuals or organizations with varying demands or needs~\cite{baruah1996proportionate,ghodsi2011dominant,joe2013multiresource,lee2019webuildai}. 
Metrics to measure allocation fairness include Jain's index~\cite{jain1984quantitative}, Max-min/min-max~\cite{radunovic2007unified}, proportional fairness~\cite{kelly1997charging}, and entropy~\cite{renyi1961measures}. Other popular metrics measure inequality as opposed to fairness, especially to quantify income inequality.
%
For example, the Gini coefficient~\cite{lorenz1905methods} and Hoover index~\cite{hoover1936measurement} determine inequality by measuring the deviation of the income distribution of the Lorenz curve from the diagonal line of perfect equality~\cite{lorenz1905methods}.
These metrics provide a single measure to indicate the inequality in a population, but have limited interpretability to discern whether the inequality comes from high or low income groups.
The Atkinson index~\cite{atkinson1970measurement} includes an inequality aversion parameter $\varepsilon$ that can be used to increase sensitivity towards changes in low incomes. 
However, for populations with diverse sub-groups, this metric does not explain which groups are most or least disadvantaged. 
This requires a metric to be \textit{additively decomposable} to attribute inequality to specific groups.
One such metric is the Generalized Entropy (GE) index~\cite{shorrocks1980class,speicher2018unified}, derived from information theory to measure redundancy. It is often used to measure the diversity of incomes. 
It also has a sensitivity parameter $\alpha$ to adjust sensitivity towards low or high incomes.
Special cases of the GE index are the mean log deviation ($\alpha=0$), Theil index ($\alpha=1$), and half of the squared coefficient of variation ($\alpha=2$).
In this work, we leverage the additive decomposibility of the GE index to satisfy a common axiom in explainability~\cite{datta2016algorithmic} to attribute inequality to specific groups with between-group and within-group inequalities.
Hence, from these decomposed inequalities, the distribution of benefits can be compared for similarity across different groups.

\subsubsection{Fair allocation in urban planning}

The fair access of scarce public resources, such as parks, hospitals and schools, is a paramount goal in urban planning.
Urban and transportation planners typically examine the accessibility to these amenities by the distance from housing to the amenity. Accessibility differs by transport modes, including walking~\cite{reyes2014walking} and public transit~\cite{foth2013towards}, and can be calculated using various metrics~\cite{neutens2010equity}.
While visualizing accessibility can indicate if benefits are unequally distributed, it is difficult to see which regions have more inequality. Hence, several methods have proposed measuring various inequality metrics, such as the Gini coefficient~\cite{cheng2021framework}, and local indicators of spatial autocorrelation (LISA)~\cite{anselin1995local}. However, these inequality metrics are typically reported for each city globally, rather than by sub-regions, though some recent works visualize them geospatially (e.g., \cite{cheng2021framework}).
In this work, we visualize housing and amenity locations, and the accessibility and local inequality of each housing location.

\subsection{Visualization and Analytical Tools}

\subsubsection{Visualization for algorithmic fairness}

Many visualization tools have been developed to examine fairness or bias in machine learning. 
Commercially available ones include IBM AI Fairness 360~\cite{bellamy2018ai}, Google Tensorflow Fairness Indicators\footnote{https://www.tensorflow.org/tfx/guide/fairness\_indicators}, and Microsoft Fairlearn~\cite{bird2020fairlearn}; these provide basic bar charts and scatter plots to show predictive parity in datasets and model predictions.
More sophisticated visualization interfaces help to examine other forms of unfairness, such as intersectional bias in FairVis~\cite{cabrera2019fairvis}, and causal fairness in Silva~\cite{yan2020silva}.
However, these only provide a high-level, global view of the model fairness, and do not identify which subgroups are particularly unfairly treated. DiscriLens\cite{wang2020visual} can identify itemsets that are discriminatory using rule mining, though these data-driven sets may not align with domain-specific groups and be less interpretable.
FairSight~\cite{ahn2019fairsight} can examine individual and group fairness, and explain the influence of each feature based on feature perturbation. This explainability method is common for black boxes, but suffers from approximation errors.
In this work, we visualize inequality across subgroups in a domain-specific visualization for urban planning, and explain the sources of inequality based on the additive decomposability of the Generalized Entropy Index, which is a white box method that is deterministically calculated. Though, we compute feature attributions for our fairness recommendation system using Shapley values, which is similar to perturbation.

\subsubsection{Visualization for fair resource allocation}

Unlike research on visualizing unfairness in datasets and predictive models, visualizations of fair resource allocation is sparser.
AlgoCrowd~\cite{yu2019fair} visualized the AI-driven matching between workers and tasks in a dashboard, showing the distributions of worker reputation and productivity, the Jain fairness index over rounds of task allocation, and an argumentation-based explanation of why each worker was allocated.
VisMatchmaker~\cite{law2016vismatchmaker} supports the comparison of trade-offs between two solutions for a job allocation task using novel but unfamiliar visualizations (number lines and glyphs).
To promote adoption, it is preferable to use visualizations that are familiar to each domain.
Talen visualized the distribution of accessibility using geographical maps that are familiar to urban planners~\cite{talen1998visualizing}.
Similar to \cite{law2016vismatchmaker}, FairVizARD~\cite{kumar2021fairvizard} enables comparing the outcomes between two matching algorithms for ride-sharing, visualizing with a map view of taxi and request locations, and a graph view showing the time series variation of several indicators.
In IF-City, we leverage well-known geographical map, heatmap, and bar chart visualizations to convey information about inequality, benefits, and accessibility.
Furthermore, while the aforementioned visualizations support fairness assessments, they do not directly support mitigating unfairness through redesigns, which IF-City does.

\subsubsection{Visualizations for Model Explainability}

Other than being fair, intelligent systems need to be understandable, thus there has been significant recent research on explainable AI (XAI)~\cite{abdul2018trends, guidotti2018survey}. 
Drawing from psychology and philosophy, Miller \cite{miller2019explanation} argued that explanations should be causally attributional, contrastive, and counterfactual. These correspond to explanations of feature attributes~\cite{lundberg2017unified, ribeiro2016should}, contrastive explanations~\cite{dhurandhar2018explanations,
miller2021contrastive}, and counterfactual explanations~\cite{cheng2020dece,gomez2020vice,wachter2017counterfactual,wexler2019if}.
Other explanations leverage sophisticated visualizations, such as partial dependence plots~\cite{abdul2020cogam,krause2016interacting}, network activations~\cite{kahng2017cti,wang2020cnn}, network summaries~\cite{hohman2019s}, saliency maps~\cite{simonyan2014deep}, feature visualizations~\cite{olah2017feature}.
The large variety of XAI techniques makes it hard for developers to prototype with them. ExplAIner~\cite{spinner2019explainer} proposed a framework to unify how XAI techniques are used, and defined a pipeline to understand, diagnose and refine models.

Some research has focused on defining workflows and developing rich dashboards combining multiple visualizations for explanations.
Krause et al.~\cite{krause2017workflow} visualized the model performance on groups of instances with similar feature importance.
To make sense of the high number of features or patterns in models, matrix layouts have been used to visualize extracted rules with RuleMatrix~\cite{ming2018rulematrix} and important features in random forests with ExMatrix~\cite{neto2020explainable}.
Supporting interactivity by investigating counterfactual cases has been particularly popular.
Prospector~\cite{krause2016interacting} visualizes with partial dependence plots and sliders, allowing users to investigate how predictions change with different feature values.
Gamut~\cite{hohman2019gamut} provides similar capabilities for generalized additive models (GAM).
DECE~\cite{cheng2020dece} provides an interactive visualization to refine counterfactual queries for subgroups of instances.
The What-If Tool~\cite{wexler2019if} provides visualizations of confusion matrices, scatter plots, histograms for iterative counterfactual explorations.
These tools were designed primarily for data scientists and engineers familiar with statistical concepts and graphs. For domain experts or lay users, simpler interactions will be more accessible.
To support the user-centered design of explanations, several frameworks have been proposed based on the human reasoning process~\cite{wang2019designing}, and patterns of user in
inquiry~\cite{liao2020questioning,lim2009assessing,lim2010toolkit}.
In this work, we implemented Lim and Dey's intelligibility taxonomy~\cite{lim2009assessing} of question types to understand and improve fairness. 
While some XAI tools can be used to investigate AI fairness (e.g., with the What-If Tool~\cite{wexler2019if}), they only support viewing predictive parity or equality~\cite{Verma2018}, and not fair allocation, which is our focus in this work.

\subsubsection{Visualization for urban analysis and planning}

To understand complex urban environments, many visualizations have been developed to view spatio-temporal movement patterns~\cite{adrienko2010spatial,holten2009force,hurter2012graph,liu2016smartadp,lyu2019od,zeng2019route}, transit networks~\cite{weng2020towards}, air pollution~\cite{deng2019airvis}, activities~\cite{miranda2016urban} and events~\cite{doraiswamy2014using}.
But visualizing inequality in cities has been rudimentary with locations and simple metrics on maps~\cite{cheng2021framework,talen1998visualizing}.

Urban planning requires more sophisticated interactive visualizations to design the urban layout, simulate stakeholder and resident behaviors in the urban environment, and visualize various urban indicators (metrics)~\cite{waddell2002urbansim}. 
Popular tools used by professional urban planners, such as ArcGIS~\cite{law2015getting}
and UrbanSim~\cite{waddell2002urbansim}, use 2D and 3D maps as the main view. Recently, tools have been designed to inform and elicit feedback from collaborators and stakeholders~\cite{mueller2018citizen,zhang2017citymatrix}. These tools calculate urban indicators to visualize geospatially, and can provide AI suggestions to optimize planning. 
However, they currently do not allow planners to assess, analyze, and mitigate inequality.
In this work, we extend the visual paradigm of urban planning tools to support these new capabilities.

%% file: ConceptualPipeline.tex

\section{Intelligible Fair Allocation Framework}

\begin{figure}[h]
\vspace{-0.1in}
\centering
\includegraphics[width=0.45\textwidth]{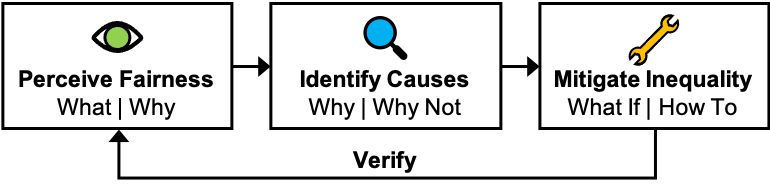}
\vspace{-0.05in}
\captionof{figure}{
Framework to perceive fairness, identify causes, and mitigate inequality with intelligibility features. 
}
\vspace{-0.05in}
\label{fig:ConceptualFramework}
\end{figure}

The scope of our work is situated in the context of computer-aided design for urban planning, where urban plans designed by urban planners are assessed with computer software to calculate various indicators (e.g., people and building density~\cite{cheng2009understanding}, accessibility and walkability~\cite{ewing2010travel}, and economic feasibility and sustainability~\cite{eu2011towards}). Specifically, we focus on 
the task of designing an urban plan to be more fair to diverse groups of residents.
This can be treated as a fair allocation problem where building resources can be automatically allocated to fairly distribute benefits to different residents.
However, due to the complexity of urban planning to balance many criteria, it is critical involve domain experts to iteratively design the urban plan.

We define fair planning as a 3-step process (Fig.~\ref{fig:ConceptualFramework}) to perceive the fairness of the design, identify causes for the unfairness, and mitigate the inequality by identifying opportunities for change.
This repeats with verifying whether fairness increased, and iterating as necessary.
Supporting these steps requires providing deeper insights by answering diverse questions that the designer may ask.
To aid non-technical users to understand how fairness is calculated and how to improve it, we leveraged the user-friendly intelligibility taxonomy of Lim and Dey \cite{lim2009assessing}. 
We propose the Intelligible Fair Allocation (IF-Alloc) Framework to provide different explanations for the measurement, examination, and mitigation of inequality in fair allocation. 
We contextualize specific intelligibility questions \cite{lim2009assessing} and corresponding explanations \cite{miller2019explanation} for fair allocation as follows:

\begin{itemize} [leftmargin=0.16in]
\item What (Status, Comparison). Indicate the prediction outcome of the system. For fair allocation, it is an explicitly calculated as a fairness or inequality indicator. Additionally, provide a historical timeline of outcomes to allow comparison across iterations to verify improvements.
\item Why (Trace, Attribution). Provide detailed traces or attributions to internal calculations for users to identify the source of a decision and their influence.
\item Why Not (Contrastive). Provide for comparison between details of two states or allocation plans. 
Comparing details across multiple iterations allows the user to track whether the system is improving over time.
\item What If (Counterfactual). Provide simulations for users to propose solutions and examine system behavior. 
Supporting this interactively can enable the user to rapidly iterate designs by tightening the feedback loop.
\item How To (Counterfactual). Provide recommendations of designs with optimal indicators to accelerate the user's iterative design, compared to unaided trial and error (with What If). 
The recommendations can also be constrained to narrow the search space of desired designs.
\end{itemize}

%% file: Methods.tex
\section{IF-City: Technical Approach}

\begin{figure}[t]
\centering
\includegraphics[width=0.48\textwidth]{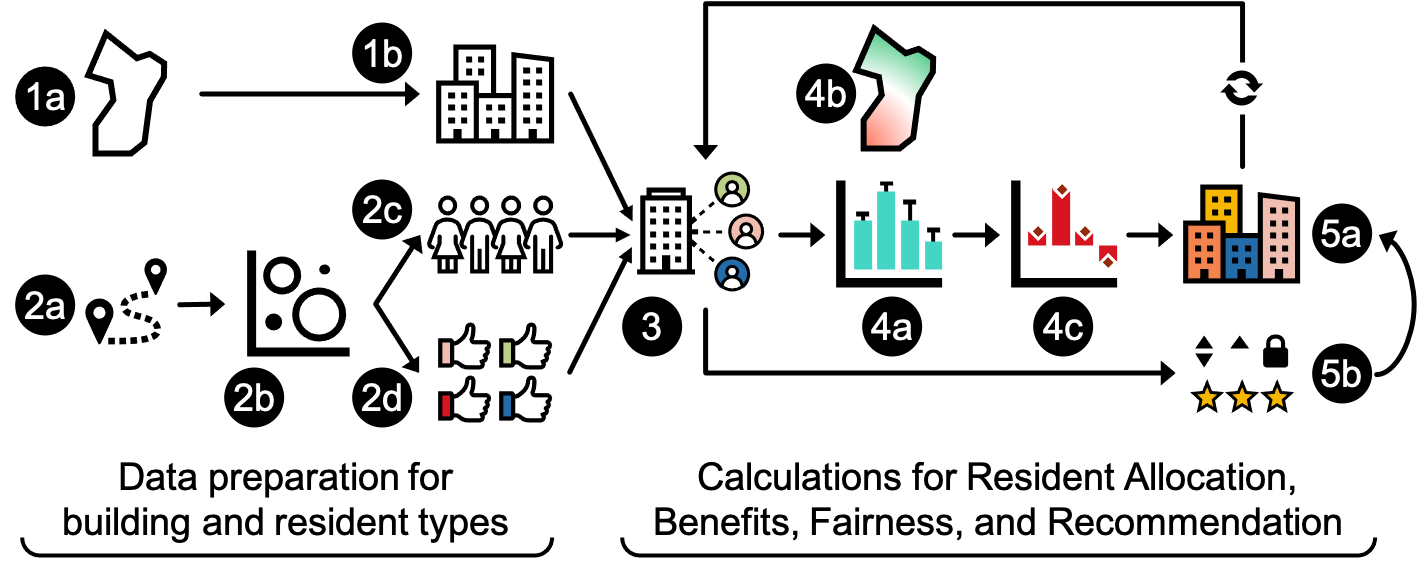}
\captionof{figure}{
System overview of data preparation and calculations.
Map and building data (1a) are used to extract building function types (1b) and their existing locations.
User trajectories (2a) are extracted from location-based social media data and clustered (2b) to identify resident types (2c) and their average preferences for different building functions (2d).
The resident allocation engine (3) simulates the migration of residents into the designed city.
The benefits for the population are calculated by resident type (4a) and location (4b), which are used to calculate the group inequality for each resident type (4c).
To support understanding the fairness indicator (4c, What), 
the user can view which resident types had unfair benefits (4a, Why) or where benefits were unfairly located (4b, Why),
redesign the urban plan to change benefits and reduce inequalities through direct manipulation (5a, What If) 
or run the recommendation engine to receive fairer design proposals (5b, How To).
}
\label{fig:SystemOverview}
\end{figure}

We describe our technical approach (Fig.~\ref{fig:SystemOverview}) to implement IF-Alloc for fair urban planning in the Intelligible Fair City Planner (IF-City). 
Specifically, we calculate:
1) the accessibility to buildings of various amenities, 
2) the benefit to each resident based on their preferences for different amenities and their accessibility,
3) the inequality based on inputting the benefits into the Generalized Entropy Index.
The inequality is additively decomposable, so this allows us to divide inequality among groups of residents by resident types. 
Computing these indicators first depends on a predetermined building layout, then the subsequent allocation of residents, which we describe in Subsection \ref{sec:what-if-simulation}.
Finally, we describe our recommendation system for fairer edits to accelerate the design iteration
These technical features respectively support the intelligibility queries --- Why, What If and How To --- which we detail in the next subsections.

\subsection{Accessibility to Building Type}\label{sec:Accessibility}

In urban planning, buildings are defined by their functions or amenities. Residential buildings provide homes for people, commercial buildings support commercial activity like retail and restaurants, educational buildings include schools and libraries, cultural buildings promote culture in museums and theaters, and parks support outdoor recreation. We define these as building \textbf{function types} $f$.

\textbf{Accessibility} is the ease of reaching spatially dispersed resources with a certain function. 
It depends on the distance to and amount of each resource.
The farther a building is from the resident, the lower the accessibility. The larger the floor area of the function in the building, the higher the accessibility
(e.g., adding offices will raise the likelihood of having a job nearby).
We compute accessibility with the gravity model \cite{paez2012measuring} 
which treats resources farther away as having inversely smaller effects.
For a specific function type $f$, the accessibility at a residential building \textit{location} $l$ is:
\begin{equation}
a_{l, f} = \sum\nolimits_{\ell \in L} \frac{v_{\ell,f}}{\rho_f} e^{-\kappa_f d_{l,\ell}},
\label{Eq: accessibility}
\end{equation}
where $\ell \in L$ is a nearby building,
$v_{\ell,f}$ is its floor area for function type $f$, 
$d_{l,\ell}$ is the distance of $\ell$ to residential building $l$,
$\kappa_f$ is the impedance to travel for the function $f$ \cite{iacono2008access}, and
$\rho_f$ is the planning priority weight for function type $f$ to prioritize for equity.
For example, residents typically need more space for a park than for retail for similar utilities, so the priority weight of parks should be higher.

\subsection{Resident Preference, Utility, and Benefit}\label{sec:Utility}

\begin{figure}[t]
\centering
\includegraphics[width=0.48\textwidth]{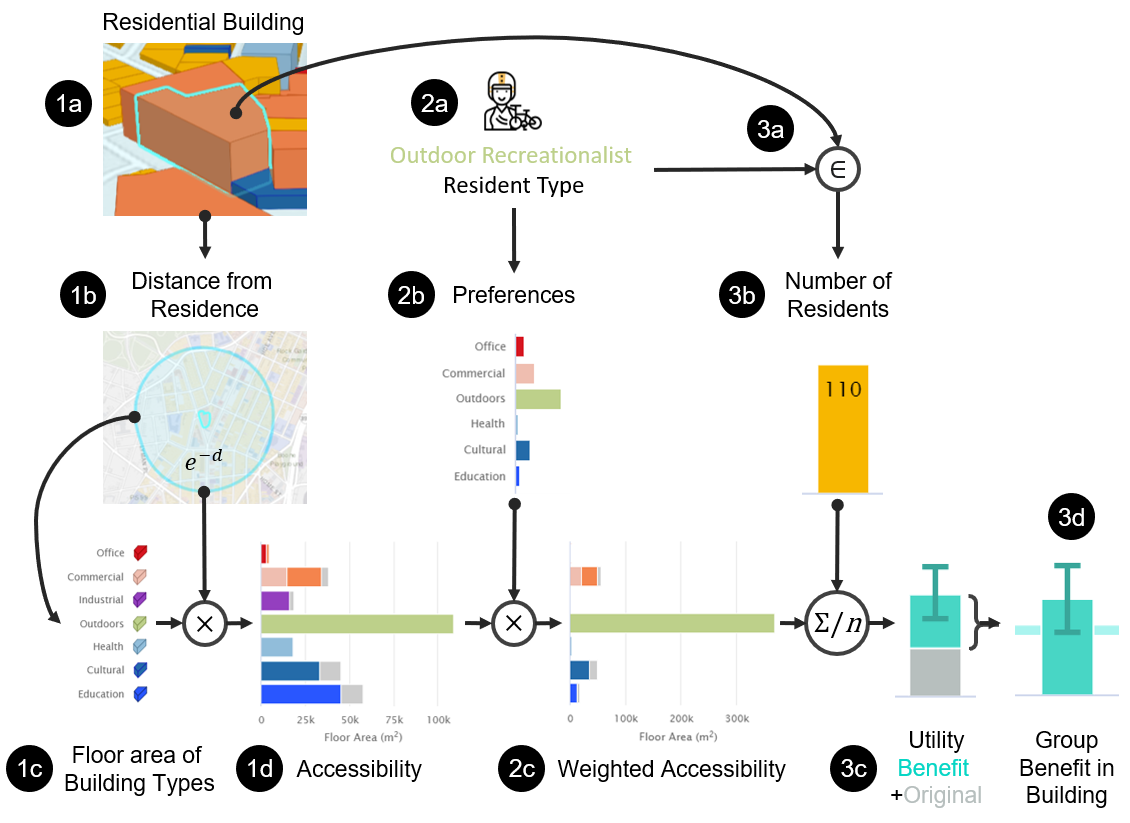}
\captionof{figure}{
Group benefit calculation for one example resident type.
For a residential building (1a), within a distance threshold radius (1b), measure the floor area of each building type (1c), and multiply by decaying distance to calculate the accessibility (1d).
For a resident type (2a), multiply their preferences (2b) with accessibility to get the weighted accessibility (2c).
For residents of the type in the residential building (3a), for each person (3b), calculate the mean weighted accessibility to get the total utility (3c).
The difference from the original utility is the benefit (3d).
}
\label{fig:BenefitCalculation}
\end{figure}

Cities attract diverse residents who have various interests or \textbf{preferences}. Even in the same residential building, different residents will value their location differently, i.e., they will have varying \textbf{utilities} at their location due to differing preferences to nearby amenities.
For example, placing a large park near a residential building will be more appreciated by residents who love outdoor activities than those who prefer cultural facilities.
We calculate the utility of individual resident $i$ living at residential building location $l$ as the preference-weighted sum of accessibility, i.e., 
\begin{equation}
u_{i,l} = \sum\nolimits_{f \in F} \pi_{i,f} a_{l,f},
\label{Eq: utility}
\end{equation}
where $F$ is the set of non-residential building function types, 
and $\pi_{i,f}$ is the resident's preference for each function type $f$.
The preferences can be acquired subjectively using preference elicitation surveys\cite{rambonilaza2007land}, or implicitly with objective data-driven measures from their visit trajectories in location-based social media \cite{bao2012location}. In IF-City, we obtained preferences with the latter approach (details in Appendix A).

Since residents would have had a prior residence before being placed in the new urban design, they would have a prior utility $u_i^{(0)}$ that is non-zero. 
This can be estimated from location-based activity data.
Hence, we seek to balance the change in utility, which we denote as \textbf{benefit}, i.e., 

\begin{equation}
b_{i,l}= u_{i,l} - u_i^{(0)}.
\label{Eq: benefit}
\end{equation}
%
We assume that residents would not relocate to the new urban design if their benefit was negative. 
Hence, allocations will not be made for such cases, and we can balance the benefits instead new utility scores (which may be negative).

Fig. \ref{fig:BenefitCalculation} summarizes the steps to calculate resident benefit using accessibility and preferences. With this definition of benefit, we can next define a fairness score for all residents for an urban design. These steps provide a Why trace explanation to identify the source of fairness or inequality.

\subsection{Generalized Entropy for Fairness Attribution (Why)}\label{sec:GE}

\begin{figure}[t]
\centering
\includegraphics[width=0.48\textwidth]{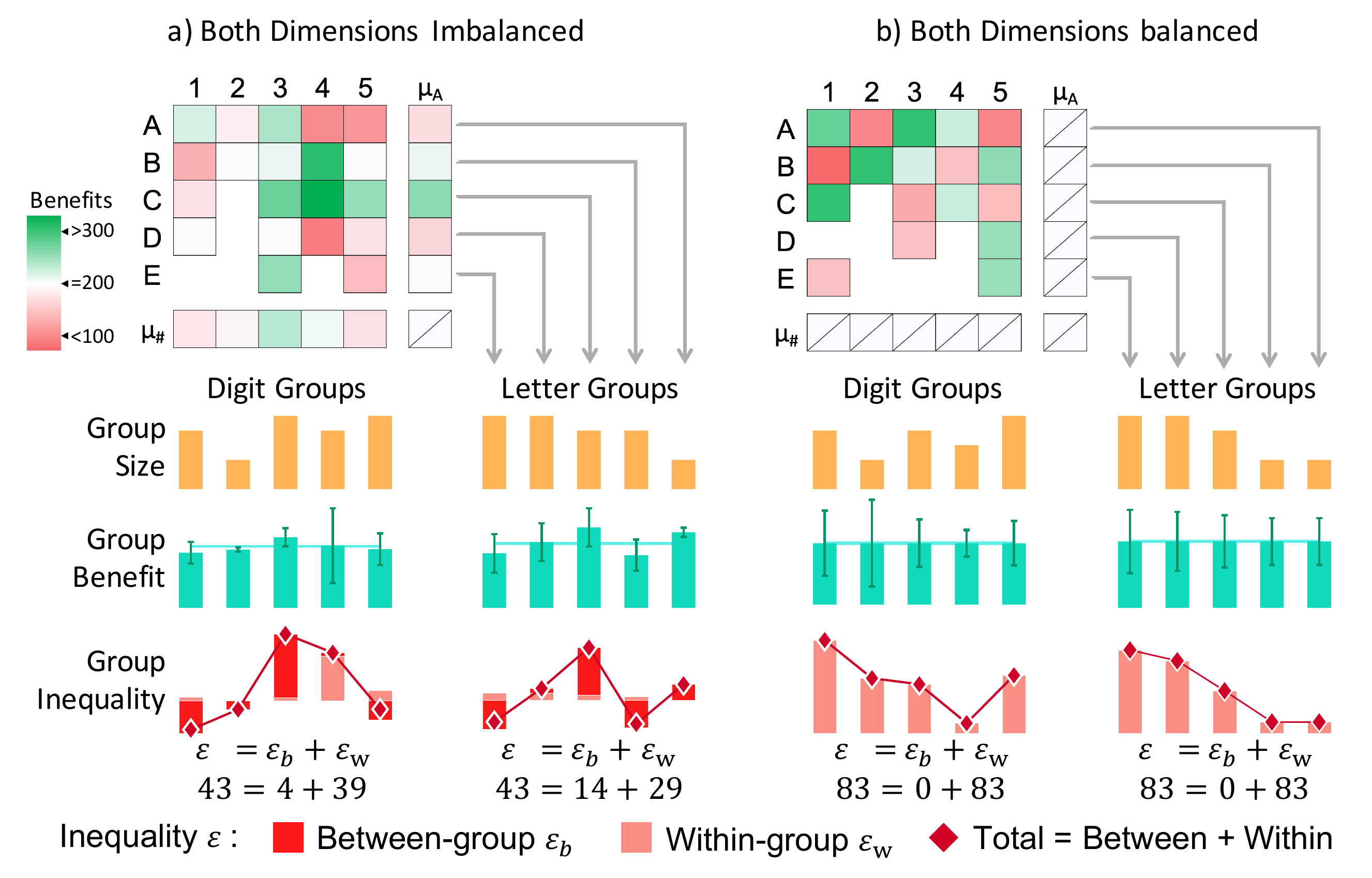}
\captionof{figure}{
Conceptual illustration of Benefit and Inequality charts. 
$5\times5$ matrices show benefits across individuals arranged along two dimensions of groups --- digits (columns) and letters (rows); bottom row and right column show group averages.
Each filled cell represents an individual with benefit value: white (average), green (above), red (below).
For example, individual A1 belongs to the Digit 1 and Letter A groups and has above-average benefit.
Yellow graphs show group sizes.
Teal graphs show group mean benefits and standard deviation; horizontal line is the global mean.
Red graphs show group inequalities: between-group (dark red), within-group (pink), and total (line).
Two examples shown: a) both dimensions imbalanced, and b) both balanced.
Interpretation: Example (b) has perfectly balanced groups in both dimensions, yet can have higher inequality than Example (a), due to high within-group variability.
}
\vspace{-0.07in}
\label{fig:ConceptualDiagramForInequality}
\end{figure}

\begin{figure}[t]
\centering
\includegraphics[width=0.48\textwidth]{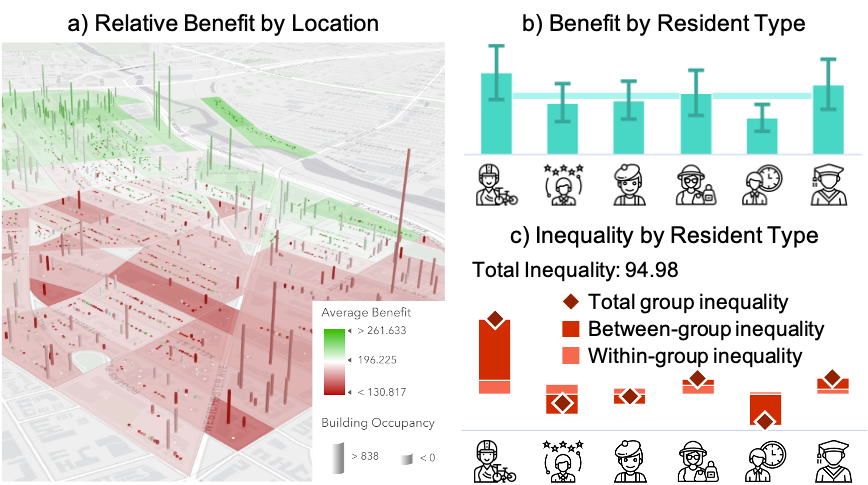}
\captionof{figure}{
Benefit and inequality visualizations in IF-City by location (a) and resident type (b, c). 
a) Relative benefit by location shown as a heatmap on a geographical map, where colored areas indicate relative benefit for each city block and cylinders for each building. White color indicates mean benefit.
b) Benefit by resident type showing group benefit mean (bars) and standard deviation (error bar) and global mean (horizontal line).
c) Inequality by resident type calculated as Generalized Entropy Index decomposable into between- and within-group inequalities per group.
}
\vspace{-0.07in}
\label{fig:BenefitInequality}
\end{figure}

We employ the Generalized Entropy (GE) index~\cite{speicher2018unified} to calculate the extent that the residents' benefits are unevenly distributed. 
It is popularly used to measure income inequality and is the generalization of information entropy~\cite{shorrocks1980class,speicher2018unified}. 
The GE index of the set of benefits $B$ of all residents is 
\begin{equation}
\varepsilon^\alpha(B) = \frac{1}{n\alpha(\alpha-1)} \sum_{i=1}^{n}\left[\left(\frac{b_i}{\bar{b}}\right)^\alpha - 1\right],
\label{Eq:GE}
\end{equation}
where 
$b_i$ is the benefit of individual resident $i$,
$\bar{b}$ is the global mean benefit,
$n$ is the number of residents (population size), and
$\alpha$ is a sensitivity parameter, $\alpha \in \mathbb{R}$. 
We choose $\alpha=2$, for $\varepsilon^\alpha$ to be more sensitive towards individuals with smaller benefits (i.e., disadvantaged groups).
%
The GE index has nice explainable properties of being additively decomposable. 

\textbf{Additive}. The GE index is a simple sum of terms for each individual's benefit $b_i$, so decomposing it into all individuals can identify which individuals have above or below average benefit, and hence affect inequality.

\textbf{Subgroup decomposability}. By partially aggregating GE index inequalities of several individual, we can calculate the subgroup inequality of non-overlapping groups $g$ within the population. To aid interpretability, these terms can be divided into between-group inequality $\varepsilon_b^\alpha(B)$ and within-group inequality $\varepsilon_w^\alpha(B)$. The total inequality is also conveniently just the sum of these components, i.e., 
\begin{equation}
\varepsilon^\alpha(B) = \varepsilon_b^\alpha(B) + \varepsilon_w^\alpha(B).
\end{equation}

The between-group inequality measures how different the mean benefits of different groups are, i.e., 
\begin{equation}
\varepsilon_b^\alpha(B) = \sum_{g \in G} \frac{n_g}{n\alpha(\alpha-1)}\left[\left(\frac{\bar{b}_g}{\bar{b}}\right)^\alpha - 1\right],
\end{equation}
where $\bar{b}_g$ is the mean group benefit, $n_g$ is the number of individuals in group $g$, and $G$ is the set of all groups. 

The within-group inequality measures how different the individual benefits within each group are, i.e.,
\begin{equation}
\varepsilon_w^\alpha(B) = \sum_{g \in G}\frac{n_g}{n}\left(\frac{\bar{b}_g}{\bar{b}}\right)^\alpha \varepsilon^\alpha(B_g),
\label{Eq:within-Inequality}
\end{equation}
where $B_g$ is the set of individual benefits of residents in group $g$. 
Fig.~\ref{fig:ConceptualDiagramForInequality} 
shows how total inequality, between- and within-group inequalities can vary with different benefit distributions and group divisions.
Fig.~\ref{fig:BenefitInequality} shows a domain-specific visualization for the benefits and group inequalities for the dimensions of resident type and locations in a city.
Inequality can be reduced by 1) lowering the average benefit of high groups, 2) raising the average benefit of low groups, or 3) reducing the variance of benefits within each group.
In IF-City, we grouped residents with similar preferences using k-means clustering on their frequency of visiting various building types (details in Appendix A).
These calculations provide a Why explanation to identify which resident types or locations raised or lowered inequality.

\subsection{Population Allocation for What If simulation} \label{sec:what-if-simulation}

The benefit and inequality calculations require the residential location $l$ of each resident $i$ to be predetermined. We determine the locations by solving a population allocation problem based on the user's urban design. Specifically, the user designs the urban plan, then the resident population is automatically allocated based on probabilities that are proportional to their benefit.
This non-trivial calculation provides the What If explanation of the user's proposal.

We first estimate the marginal probability $p_i$ of resident $i$ moving to the city. $p_i$ should be monotonic to the average benefit $\bar{b}_i = \sum_{l\in L_\text{Res}} b_{i,l} / |L_\text{Res}|$ of the resident living in all residential building locations $L_\text{Res}$ in the city, where $b_{i,l}$ is the resident's benefit of living at building $l$. 
We normalize this as a probability with a $\tanh$ transform function, i.e., 
\begin{equation}
p_i = \max (\tanh(\gamma\bar{b}_i), 0),
\label{Eq:ProbU}
\end{equation}
where $\gamma > 0$ is a normalization parameter that is determined by the constraint that the total probability of all residents should equal to the sum of occupancy of all residential buildings in the city, i.e., $\sum_{i} p_i = \sum_{l} o_l$, where $o_l$ denotes the occupancy of residential building $l$. 
Note that $\bar{b}_i$ could be negative because the resident may not benefit from moving to the city. In this case, the probability $p_i$ is 0.

Given the marginal distribution $p_i$ for the resident $i$ for the whole city, we seek to estimate probability $p_{i,l}$ of the resident moving into each specific building $l$.
We perform iterative proportional fitting (IPF) ~\cite{frick2003generating,muller2010population} to estimate $p_{i,l}$ by treating $\{p_{i,l}\}_{I,L}$ as a matrix of rows $i$ and columns $l$.
At each iteration step $\eta$, for each residential building location $l$ with vacancy, IPF estimates $p_{i,l}^{\eta}$ with the following two equations for the odd and even steps, respectively: 
%
\begin{equation}
\hat{p}_{i,l}^{(2\eta-1)} = \frac{\hat{p}_{i,l}^{(2\eta-2)} p_i^{(2\eta-2)}}{\sum_{l} \hat{p}_{i,l}^{(2\eta-2)}},
\quad
\hat{p}_{i,l}^{(2\eta)} = \frac{\hat{p}_{i,l}^{(2\eta-1)} o_l^{(2\eta-1)}}{\sum_{i} \hat{p}_{i,l}^{(2\eta-1)}},
\label{Eq:IPF_1}
\end{equation}
where $p_i^{(2\eta-2)}=\sum_{l} p_{i,l}^{(2\eta-2)}$ and $o_l^{(2\eta-1)}=\sum_{i} p_{i,l}^{(2\eta-1)}$ are the marginal distributions across individuals and locations, respectively.
The iterations terminate when $p_{i,l}$ converges to a small threshold.
The resident $i$ is thus allocated to building $l$ with the probability $p_{i,l}$.
Residents are allocated in a random order until all buildings are fully occupied.

\subsection{Inequality Mitigation Recommendation (How To)}\label{sec:HowTo}

While the allocation system (What If simulation) is a powerful tool to provide feedback on the user's design, it can be tedious for users to iterate designs on their own, leading to less productive trial-and-error.
To speed up the design iterations, we provide an inequality mitigation recommendation system for How To design for more fairness.

Our approach to generate the counterfactual explanations is distinct from current approaches in machine learning by:
1) recommending a numeric outcome and optimizing its value, rather than an alternative target label for categorical classification~\cite{wachter2017counterfactual};
2) supporting user-constrained recommendations by solving a constraint optimization problem, rather than using a slow brute-force search method~\cite{gomez2020vice};
3) recommending in terms of coarser user-manageable features, rather than raw fine-grained features;
and
4) indicating the importance of each change by attributing their contribution towards the improved outcome using Shapley value calculations (inspired by SHAP~\cite{lundberg2017unified}).

\subsubsection{Problem Formulation}
An urban design is the set of floor areas for all buildings of different function types in the city.
Given the current design, the user's goal is to propose a design change that decreases the inequality index.
Instead of recommending changes to specific buildings, we recommend changes to a coarser unit of geography --- the census block $k \in K$. This reduces the search complexity for an optimal solution, aligns with planning practice to consider planning at coarser granularity, and allows for more freedom in designing specific buildings.

Let $\delta_{k,f} = \Delta v_{k,f}$ denote the floor area change of census block $k$ and building function type $f$, $v_{k,f}$ denotes the floor area of existing block $k$ and type $f$. The new design $v_{k,f} + \delta_{k,f}$ determines the allocation of residents (i.e., $p_{i,l}$) and thus a distribution of benefits $B$ for all the allocated residents. 
We seek the optimal $\delta_{k,f}$, such that the inequality index $\varepsilon^\alpha(B)$ defined by Eq.~\ref{Eq:GE} is minimized, i.e.,
\begin{align}
& {\text{minimize}}
& & \varepsilon^\alpha(B) &\label{Eq: obj}\\
& \text{subject to}
& & \delta_{k,f} + v_{k,f} \geq 0, & \label{Eq: C1}\\
& & & \sum\nolimits_{f} \left(\delta_{k,f} + v_{k,f}\right)/s_k - \bar{h}_k \leq h_\Delta^{(\text{max})}, & \label{Eq: C2}\\
& & & \sum\nolimits_{k} \sum\nolimits_{f} |\delta_{k,f}| \leq \delta^{(\text{max})}, & \label{Eq: C3} \\
& & & |\sum\nolimits_{k} \delta_{k,f=\text{Res}}| \leq \delta_{f=\text{Res}}^{(\text{max})}. & \label{Eq: C4}
\end{align}
%
Eq.~\ref{Eq: C1} constrains that the final floor area cannot be negative, since that is physically impossible.
Eq.~\ref{Eq: C2} constrains the increase in average height to less than a small threshold $h_\Delta^{(\text{max})}$, where $s_k$ is the total building footprint (ground floor area) in the block $k$ and $\bar{h}_k$ is a current average height in the block; this preserves the skyline of blocks.
Eq.~\ref{Eq: C3} constrains how much the floor area in the whole city can change with a construction budget $\delta^{(\text{max})}$.
Eq.~\ref{Eq: C4} constrains the total change of residential floor area to be smaller than a threshold $\delta_{f=\text{Res}}^{(\text{max})}$ to stabilize population changes. 

Furthermore, as urban planners may want to control the increase or decrease of group mean benefits for some resident types, we add an objective to constrain the generated design recommendations.
%
Given any arbitrary new design, let $\Delta \bar{b}_{g^+}$, $\Delta \bar{b}_{g^-}$ and $\Delta \bar{b}_{g^0}$ denote the group average benefit difference between the new design and the current design for the groups whose mean benefits are to be increased ($g^+ \in G^+$), decreased ($G^-$), and kept unchanged ($G^0$), respectively. The objective is to limit  $\Delta \bar{b}_{g^+} > 0$, $\Delta \bar{b}_{g^-} < 0$, and $|\Delta \bar{b}_{g^0}| \approx 0$. 
We define an solution penalty as
\begin{align}
 \phi(G^+, G^-, G^0) & = \sum\nolimits_{g^+ \in G^+} - \min(\Delta \bar{b}_{g^+} + \tau, 0) \\
                        & + \sum\nolimits_{g^- \in G^-} \max(\Delta \bar{b}_{g^-} - \tau, 0)\\ 
                        & + \sum\nolimits_{g^0 \in G^0} \max(|\Delta \bar{b}_{g^0}| - \tau, 0),
\end{align}
where $\tau$ is a small threshold with $\tau \geq 0$. 
Adding this penalty function to Eq.~\ref{Eq: obj} gives the new objective function
\begin{equation}
{\text{minimize}}\quad
\varepsilon^\alpha(B) + \lambda \phi(G^+, G^-, G^0),\label{Eq: obj2}
\end{equation}
which searches for $\delta_{k,f}$ that minimizes inequality while satisfying the design constraints. $\lambda$ is a penalty hyperparameter, 
which we calibrated to be large, i.e., $\lambda \gg 1$.
\subsubsection{Heuristic Solution}

Since finding the optimal $\delta_{k,f}$ can be inefficient, we employ the Frank-Wolfe conditional gradient method~\cite{jaggi2013revisiting} to iteratively estimate its value by linearly approximating the objective function in Eq.~\ref{Eq: obj2}, which we denote as $m(\bm\delta + \bm v)$,
where $\bm{v} = \{v_{k,f}\}_{k\in K, f\in F}$ is a matrix denoting the floor areas of each function type $f$ in each census block $k$ and 
$\bm{\delta} = \{{\delta}_{k,f}\}_{k\in K, f\in F}$ is a matrix denoting the changes in floor areas to recommend. 
%
Algorithm~\ref{alg:FrankWolfe} starts with the current design
$\bm{v}$ that is within the search space $\bm P$ defined by the constraints (i.e., Eqs.~\ref{Eq: C1} to \ref{Eq: C4}). 
Note that this is slightly different from the standard approach to start at an arbitrary point.
In each iteration $c$, it approximates the objective function $m(\bm{\delta}+\bm{v})$ with the first-order Taylor series expansion about the point $\bm{\delta}_c+\bm{v}$, i.e., $m(\bm{\delta}_c+\bm{v}) + \nabla_{\bm{\delta}} m(\bm{\delta}_c + \bm{v})^\top (\bm{\delta} - \bm{\delta}_c)$. 
Minimizing this linear equation finds $\bm{\delta}$ that points in the direction towards the optimum. 
We then update $\bm{\delta}_{c+1} = \bm{\delta}_c + \zeta (\bm{\delta} - \bm{\delta}_c)$, where $\zeta = \frac{2}{c+2}$ is the step size. 
The algorithm terminates when the difference between consecutive objective values is smaller than a threshold, i.e., $m_{c+1} - m_c < \epsilon$. 

\begin{algorithm}[!t]
\small
\renewcommand{\algorithmicrequire}{\textbf{Input:}}
\renewcommand\algorithmicensure {\textbf{Output:} }
\caption{Frank–Wolfe Algorithm} \label{alg:FrankWolfe}
\begin{algorithmic}[1]
\State \textbf{Input:} Current design $\bm{v} = \{v_{k, f}\}_{K,F}$
\State Initialize $c = 0$, $\bm{\delta}_0 = \{0\}_{K,F}$
\While {$\Delta m > \epsilon$}
\State $\bm{\delta} = \argmin_{(\bm{\delta}+\bm{v}) \in \bm{P}} \bm{\nabla_\delta} m(\bm{\delta}_c + \bm{v})^\top (\bm{\delta} - \bm{\delta}_c)$  
\State $\bm{\delta}_{c+1} = \bm{\delta}_c + \zeta (\bm{\delta} - \bm{\delta}_c)$, $\zeta = \frac{2}{2 + c}$
\State $m_{c+1} = m(\bm{\delta}_{c+1} + \bm{v})$
\State $\Delta m = m_{c+1} - m_c$
\State $c = c + 1$
\EndWhile
\State \textbf{Output:} $\bm{\delta}_c = \{{\Delta v}_{k, f}\}_{K,F}$

\end{algorithmic}
\end{algorithm}

\subsubsection{Fairness Attribution to Edited Blocks}

The recommendation will propose edits of multiple function types for several census blocks.
It may recommend editing many blocks, which is tedious for the user.
To help users prioritize which blocks to edit, we calculate another attribution explanation to indicate the partial contributions of each block towards improving the overall fairness. Inspired by SHAP \cite{lundberg2017unified} for explaining machine learning classifiers, we calculate these attributions as Shapley values
~\cite{roth1988shapley}, which fairly measures the contributions independent of calculation order. 
For a block $k$ among the set of recommended blocks to edit $R$, $k \in R$,
we calculate its attribution towards decreasing inequality as
%
\begin{equation}
\frac{1}{|R|}\sum_{S \subseteq R \setminus \{k\}} 
\overbrace{-(\varepsilon^\alpha(B_R) - \varepsilon^\alpha(B_S))}^\text{contribution towards fairness} 
\biggm/ {\overbrace{C(|R|-1,|S|)}^\text{\# of combinations}}
\end{equation}
where $S$ is a subset of $R$ without block $k$; 
$B_S$ is the set of benefits after editing blocks $S$ as recommended,
%
$\varepsilon^\alpha(B_S)$ is the inequality due to benefits $B_S$;
$B_R$ and $\varepsilon^\alpha(B_R)$ are the benefits and inequality due to all recommended block edits.
%
$C(|R|-1,|S|)$ denotes the number of combinations $S$ that exclude block $k$ choosen from set $R$.
Since permuting all combinations is time consuming, we adopt a sampling technique~\cite{castro2009polynomial} that uses the average attribution of a few permutation samples to approximate the Shapley value to speed up calculations. 
Note that the attributions are only representative if all the recommended changes are executed.

\subsection{Intelligibility Features in IF-City}\label{sec:XAIFeatures}

We summarize the methods for intelligibility features to help users to understand inequality in an urban design.

\textbf{Why-Trace Explanation}.
By articulating how the inequality index is calculated from accessibility and preferences to utility and benefit, we can provide a trace of underlying calculations and metrics. By diagnosing the stage of the calculation, users can assess if inequality is due to building placement causing poor accessibility (user controllable), or due to resident preferences (less controllable).

\textbf{Why-Attribution Explanation}. 
The additive property of the Generalized Entropy (GE) index allows the inequality index to be subdivided into individual and group inequality. Thus, users can attribute whether the source of inequality is from differences between groups or variance within specific groups. We use this to communicate the inequality across resident types.
The attribution explanation supports the \textbf{Why} explanation for the current design and the \textbf{Why Not} explanation when used to compare different designs.

\textbf{What If Counterfactual Explanation}. 
Users may want to explore how an intelligent system would behave given an counterfactual instance of interest. In the case of urban planning, the instance is a hypothetical urban design. 
Calculating the system outcome (inequality) is sophisticated and requires first solving a stochastic allocation problem to determine where each resident would live to maximize his/her benefit.
The benefit and inequality distribution can then be calculated from the allocations.

\textbf{How To Counterfactual Explanation}. 
We provide a How To counterfactual explanation to recommend a draft solution for a fairer urban plan and accelerate design iterations. 
It proposes edits to floor areas of each function type for a few census blocks rather than specific buildings to support more creative freedom in design.
By attributing which blocks are more important for improving fairness, users can understand what edits to prioritize and be convinced to follow the recommendations.
Finally, users can constrain the recommended edits to retain some results of the prior design (e.g., do not decrease the benefit of a specific resident type).

%% file: VisualDesign.tex
\section{IF-City: Visual Design}\label{sec: VisualDesign}

\begin{figure}[t]
\centering
\includegraphics[width=0.485\textwidth]{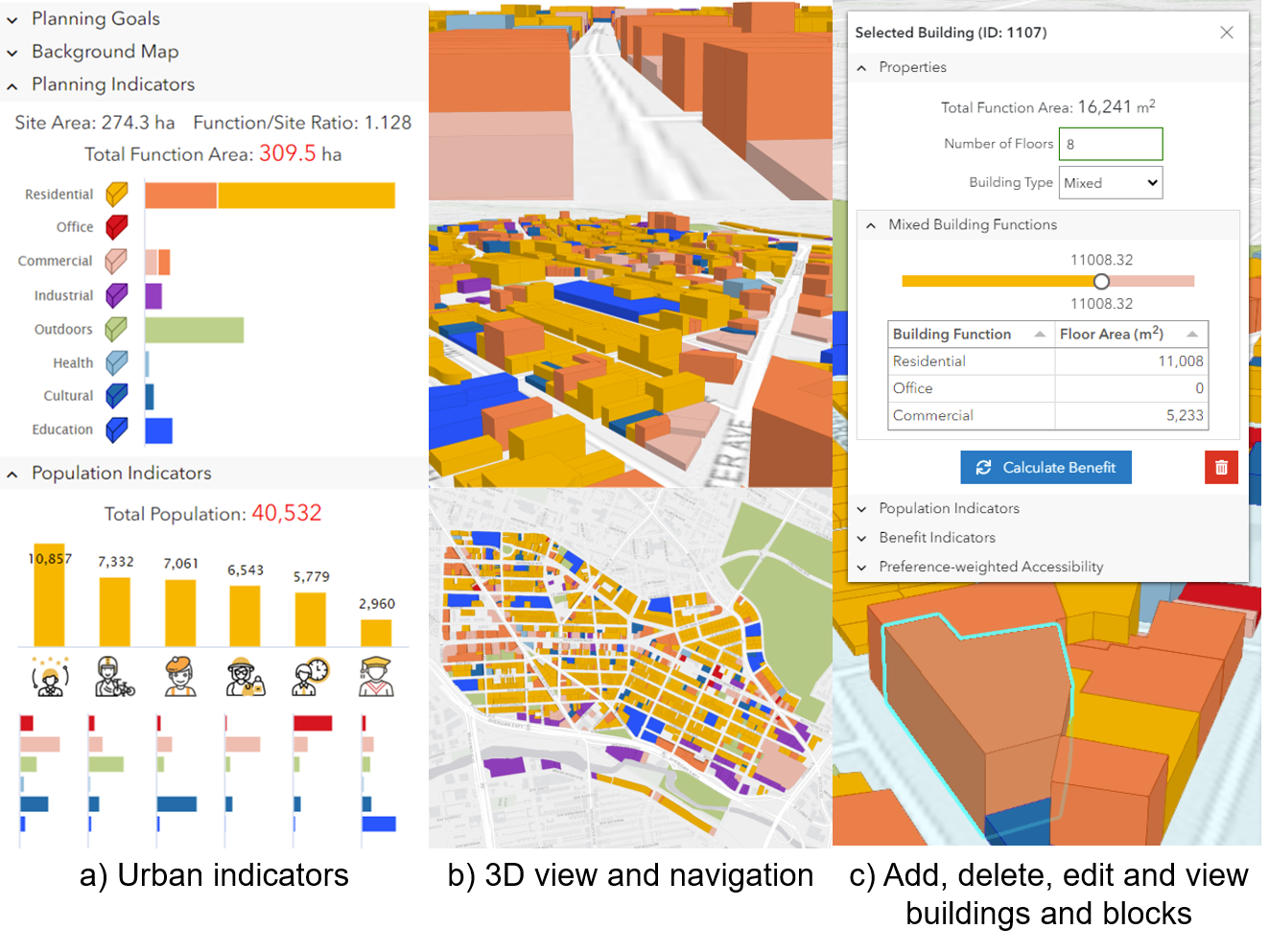}
\captionof{figure}{
Basic features IF-City for urban design. 
a) Charts showing Planning Indicators for floor areas for each function type, and Population Indicators for number of residents of each type with their respective preferences for function types. 
b) 3D map showing buildings that are color-coded by building type as indicated in the Planning Indicators. Users can navigate around the city by zooming, panning, and rotating the 3D visualization.
c) Users can view details by selecting blocks or buildings and open a pop-up dialog. This shows floor area, height, building type; for residential buildings, this further shows population and benefit indicators, and accessibility to other building types. Users can edit the urban design by adding, editing or deleting buildings.
}
\label{fig:BasicFeatures}
\end{figure}

\begin{figure*}[t]
\centering
\includegraphics[width=1\textwidth]{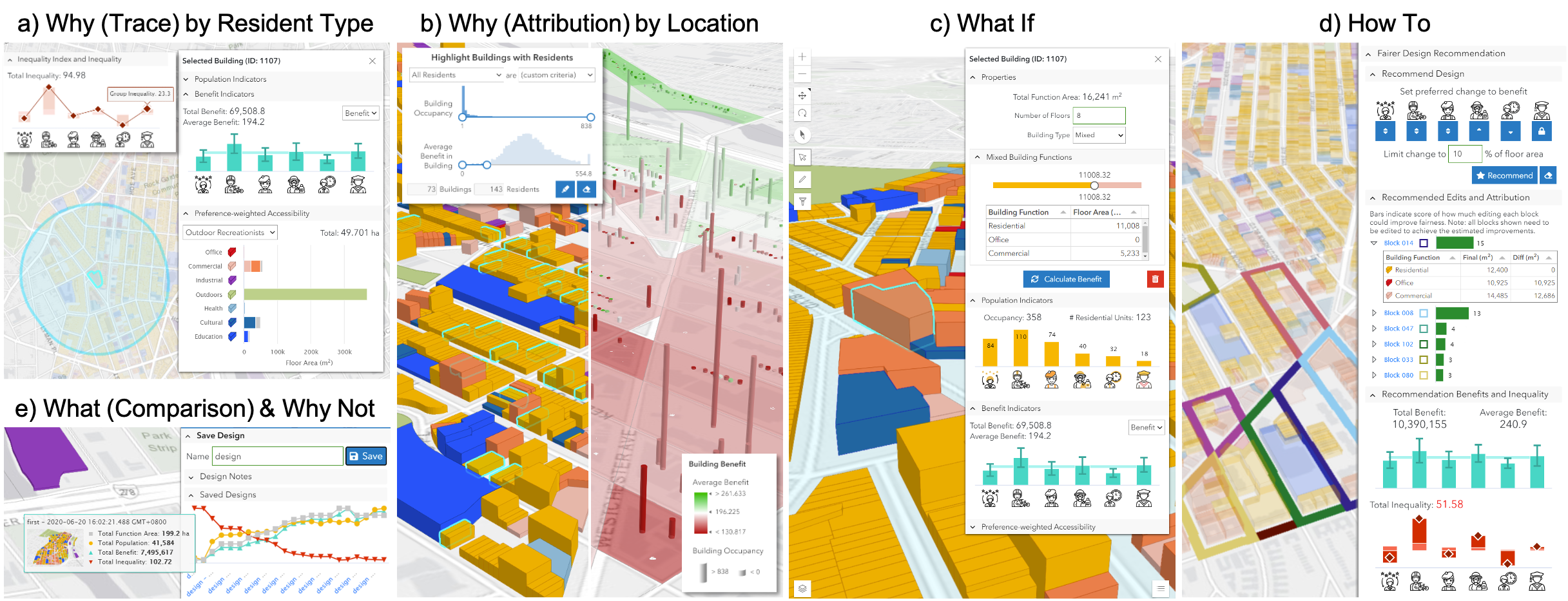}
\captionof{figure}{
Intelligibility features of IF-City that explain the fairness of the urban design.
a) Why \textit{by Resident Type}: View benefits of each resident type for the whole city, per block and per building, and accessibility at each residential building.
b) Why \textit{by Location}:  Highlight (cyan outline) buildings filtered by occupancy and average benefit, or view relative benefit of each block and building as a heatmap 
(green, red, white for above-, below-, average, respectively; darker colors for farther from average; colored ground areas for block-level benefits, 3D cylinders for building-level benefits with height for number of occupants).
c) What If: Change urban design by drawing new buildings, deleting existing ones, or editing them. Building heights and types can be edited. Occupancy and benefits are recalculated after changes.
d) How To: Request recommendation for which blocks to edit and how much floor area to change for each function type to reduce inequality. Estimated benefits and inequalities along with 
their attribution towards fairness improvement are shown.
Users can constrain recommendations to specify whether benefits should increase, decrease, or be fixed for each resident type, and the percentage of floor areas allowed to be changed. 
e) What (Comparison) and Why Not: View timeline of design iterations showing changes in indicators. Click on a time point to open the old design for detailed comparison.
}
\label{fig:IntelligibleFairnessFeatures}
\end{figure*}

Having derived the indicators and explanations to interpret fairness in urban planning, we next describe how we convey them through visual components of IF-City. 
Fig. \ref{fig:DesignDashboard} shows the whole application dashboard and Figs. \ref{fig:BasicFeatures} and \ref{fig:IntelligibleFairnessFeatures} show basic urban design and intelligible fairness features, respectively.
Inspired by Qua-Kit~\cite{mueller2018citizen},
we developed IF-City as a web app with 3D map and colored blocks for buildings.

\subsection{Basic Features for Urban Design}

\textbf{Urban (Planning and Population) indicators.} 
Urban planners track indicators to understand various characteristics of an urban design. We present these as numbers (e.g., total population, site area) and with charts to indicate subgroup information. Planning and population indicators (Fig. \ref{fig:BasicFeatures}a) are presented on the right side bar of the dashboard (of Fig. \ref{fig:DesignDashboard}), and indicate the total floor area of each building function type and population of each resident type, respectively. 
The different preferences of each resident type are also presented as vertical Preference Charts with colors corresponding to each function type.
Tooltip label hints are also provided.

\textbf{3D map view and navigation.}
Users can examine the urban design by panning, zooming, and rotating a 3D map (see Fig. \ref{fig:BasicFeatures}b). We use 3D instead of 2D so that users can view top-down to understand the location context, perceive building shapes and heights from various angles, and view at street-level for immersion. A satellite view (top of Fig. \ref{fig:DesignDashboard}) is also provided to show the context of the neighborhood. Colors of each building corresponds to its function type. 

\textbf{Editing buildings.}
Users can edit the design by adding, changing, or deleting buildings. IF-City supports planning at the granularity of specifying building footprints, heights, and function types. This is suitable for neighborhood planning, is more fine-grained than zoning, and more coarse-grained than specifying building tenants (e.g., retail or restaurants can occupy commercial spaces). Clicking a building shows a dialog (Fig. \ref{fig:BasicFeatures}c) with floor area, number of floors, function type, and residential population 
and accessibility. Mixed building types include Residential, Office and Commercial functions, and users can set their ratios. Users can edit the details and press the "Calculate Benefits" button to rerun the resident allocation simulation and recalculate the benefits and inequality indicators; this will change the population distributions in the building and the wider city. Users can also draw new buildings or delete existing ones.

\subsection{Intelligible features for fair urban design}\label{sec:XAIVisualFeatures}

IF-City has specialized features to indicate and explain fairness in the urban design. Though central to the contributions of the paper, to ensure usability, we designed these features to blend in sensibly with the primary task of the dashboard, i.e., viewing and designing the city. Therefore, the benefit and inequality features supplement the basic urban planning features, rather than taking central focus in the UI design. We highlight these intelligibility features that occur across various parts of the dashboard (Fig. \ref{fig:IntelligibleFairnessFeatures}).

\textbf{Inequality indicators (What).} Users can perceive fairness by the Total Inequality indicator (side panel in Fig.~\ref{fig:DesignDashboard}).

\textbf{Inequality attributions (Why).}
Users can identify sources of inequality by resident type or location.
Fairness across resident types can be checked by perceiving whether levels are flat in the Benefits Chart with small error bars and whether between- and within-group indices are small in the Inequality Chart (see Fig. \ref{fig:BenefitInequality}b). 
Fairness across locations can be studied by highlighting buildings with specific occupancy and average benefit ranges using Building Filtering (Fig. \ref{fig:IntelligibleFairnessFeatures}b, Left), or checking for dark green or red blocks and cylinders (for buildings) in the Benefits Heatmap (see Figs. \ref{fig:BenefitInequality}a and \ref{fig:IntelligibleFairnessFeatures}b, Right). 

\textbf{Inequality traces (Why).}
Focusing on a building, users can trace its source of inequality by clicking it to examine its Accessibility Circle and its dialog popup (Fig. \ref{fig:IntelligibleFairnessFeatures}a). Clicking on a block will also show similar information as for buildings.
The blue Accessibility Circle shows which nearby buildings were included to calculate accessibility and utility, so users may want to edit them to improve inequality. The building-specific Benefits Chart shows benefits distribution for occupants within the building. The Accessibility Chart shows the accessibility for each function type, and has the option to show preference-weighted accessibility (utility) for each resident type. For example, in Fig. \ref{fig:IntelligibleFairnessFeatures}a, Outdoor Recreationalist residents at the selected building have very high utility due to the nearby park in the northwest, leading to above-average benefit compared to other resident types and contributing to between-group inequality.

\textbf{Resident Allocation and Recalculations (What If).}
As an extension of the editing capabilities of the baseline interface, recalculating and visualizing the benefit and inequality indicators support users to ask What If questions regarding how design changes can impact the fairness outcome (Fig. \ref{fig:IntelligibleFairnessFeatures}d).
After editing buildings and pressing ``Calculate Benefit", 
the user will see updates to population, benefit and inequality indicators per building and for the whole city.

\textbf{Fairer Redesign Recommendation (How To).}
To accelerate the design iterations, users can query for recommendations on which census blocks to edit, and how much floor area of each building function type to change (Fig. \ref{fig:IntelligibleFairnessFeatures}e). The user can then choose how they want to edit buildings in recommended census blocks. Users can constrain the recommended changes that can be proposed, such as locking the benefits of resident types to remain unchanged, restricting benefits of specific resident types to only increase or decrease, and limiting the percentage change in total floor area. The recommended blocks are highlighted with thick colored outlines on the map with corresponding bars in an attribution chart with square icons of the same outline color. 
The green attribution bar chart shows the importance of each block towards reducing inequality and ranks them by decreasing attribution.
These attributions are calculated using Shapley values, and are based on the user editing all blocks as recommended. On expanding each attribution bar, users can view a table of recommended floor area change for each function type for that block. Below the attribution chart, users can see the expected benefit and inequality if the recommendation is followed. 
After accepting a recommendation, the user needs to edit the city design and recalculate (What If) to assess the design change.

\textbf{Design iteration timeline (What (Comparison)) and saved iteration (Why Not).}
For complex iterative design tasks, it is important to provide feedback for each iteration for users to assess what is improving, and what may be traded-off. Fig. \ref{fig:IntelligibleFairnessFeatures}e shows a Timeline View of saved designs for users to track and compare planning indicators, including inequality.
Clicking on a specific time point will load that design in a new browser window, showing the full dashboard with a map and charts. Using multiple screens, users can compare between urban designs for the overall city, per building, location, or per resident type.

%% file: SystemArchitecture.tex
\section{System Implementation}

IF-City was developed as an interactive web app with HTML5 and JavaScript front-end, and Python back-end with a Flask web server and custom code for the application logic. 
The 3D map and navigation interface was built using the ArcGIS JavaScript API and the charts was implemented using HighCharts.
For the How To recommendation engines, we utilized PyTorch to speed up gradient calculations and Gurobi as the optimization solver.
Resident type clustering was performed using scikit-learn.

%% file: CaseStudy.tex

\section{Case Study: Neighborhood Re-Planning}\label{sec: CaseStudy}
We situate fair urban planning within a fuller workflow of the urban planning process with multiple stakeholders and multiple steps, as defined by McLoughlin~\cite{mcloughlin1969urban}: 

\begin{figure}[t]
\centering
\includegraphics[width=0.48\textwidth]{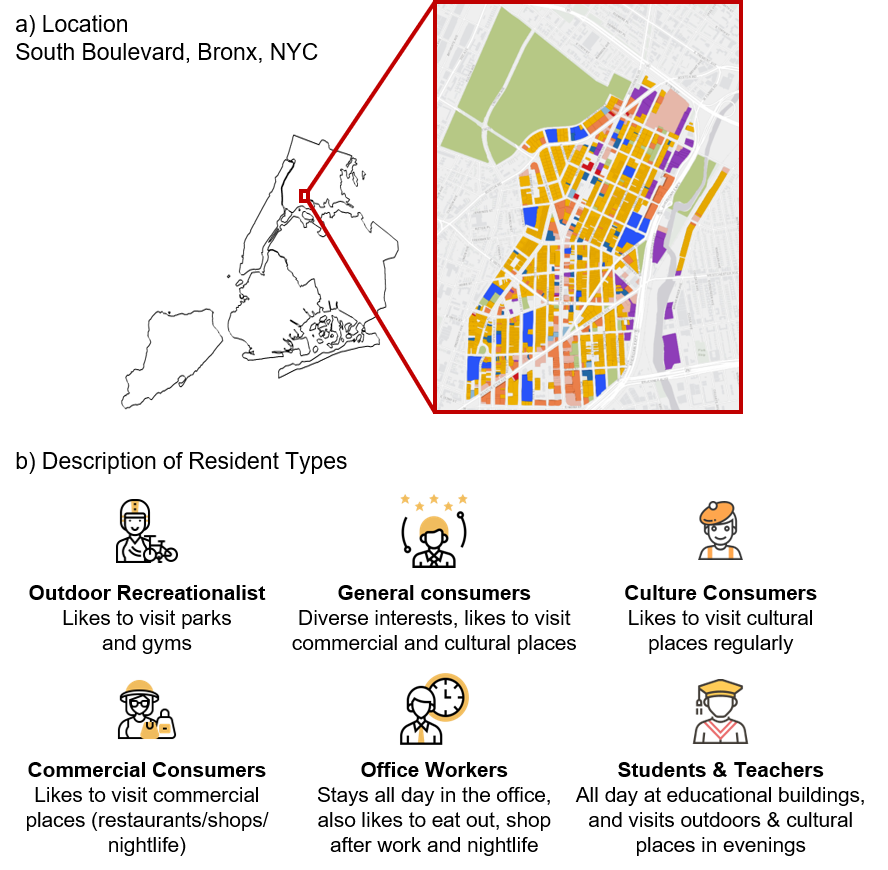}
\captionof{figure}{
Case study information. We used building and planning data regarding the South Boulevard neighborhood in the Bronx of New York City (NYC), United States. Building specifications follow the NYC  Department of City Planning Documentation, and building data is from the PLUTO dataset\cite{mappluto}. 
Six resident types with distinct preferences for different building types were derived from a cluster analysis of check-in trajectory data of 4247 NYC Foursquare users.
}
\label{fig:CasestudyBackground}
\end{figure}

\begin{enumerate} [leftmargin=0.16in,itemsep = 0in,topsep=0.01in]
\item Decide to conduct planning and choose methods; 
this can take a long time to review methods and techniques.
\item Formulate goal and identify objectives for planning by appropriate agencies, including clarifying how planning will relate to other forms of communal action. This will result in a \textit{urban design brief} to specify planning goals.
\item Study possible courses of action with the aid of models of the environment. These studies can show how the system (urban design) might behave over time due to private actions, public activities and interventions. 
\item Evaluate these courses of action in order to select an operational course by reference to assumed social values and the estimation of costs and benefits.
\item Implement the plan including direct works and continuous control of public and private proposals for change. 
\end{enumerate}
We focus on design exploration and evaluation (Steps 3-4).

\subsection{Background and Design Brief}

We present a case study of the neighborhood of Southern Boulevard in Bronx, New York City, to redesign it for fairer benefits across diverse resident types. 
Southern Boulevard (Fig.~\ref{fig:CasestudyBackground}a) is home to almost 60,000 residents. Urban planners have identified a number of gaps and opportunities to improve this neighborhood\footnote{https://www1.nyc.gov/site/planning/plans/southern-blvd/southern-blvd-updates.page}.
To examine how re-designing this neighborhood affects the re-allocation of residents, we simulated a synthetic population based on the check-in data of 4247 users from Foursquare\footnote{https://foursquare.com} collected in NYC (technical details in Appendix A). 
We identified six resident types, based on their preference for different building function types: Outdoor Recreationalists, General Consumers, Culture Consumers, Commercial Consumers, Office Workers, and Educators \& Students (Fig.~\ref{fig:CasestudyBackground}b). 
We pose a design brief with a planning goal to decrease the inequality indicator from 95 to $\leq$60 and increase the average benefit from 190 to $\geq$220, while maintaining the population and total building floor area to within 10\% of the original indicators.
This planning goal was also used later in the user study.

\textbf{Assumptions.} Since urban planning is a complex problem, in this case study, we simplified the planning process that may result in some limitations.
1)~We mined resident preferences from social media which might be biased, as it excludes many residents who are non-users. Our clustering of preferences was also based on locations that can be visited, and exclude less tangible activities.
Nevertheless, our activity-based preferences can be complemented with contemporary approaches that collect demographic-based preferences from surveys of urban residents.
2)~We allocated residents to residential buildings individually in the What If simulation, and neglected modeling families or households that move together. Further work can perform allocation at the household level with multiple occupants together.
3)~We calculated accessibility based on straight-line distances and walkability, which is a common approach. Extending this work to include transportation planning can consider transportation networks (roads, pathways, tracks) and transportation modes (e.g., cars, bicycles, trains).

\subsection{Walkthrough}

\begin{figure}[t]
\centering
\includegraphics[width=0.48\textwidth]{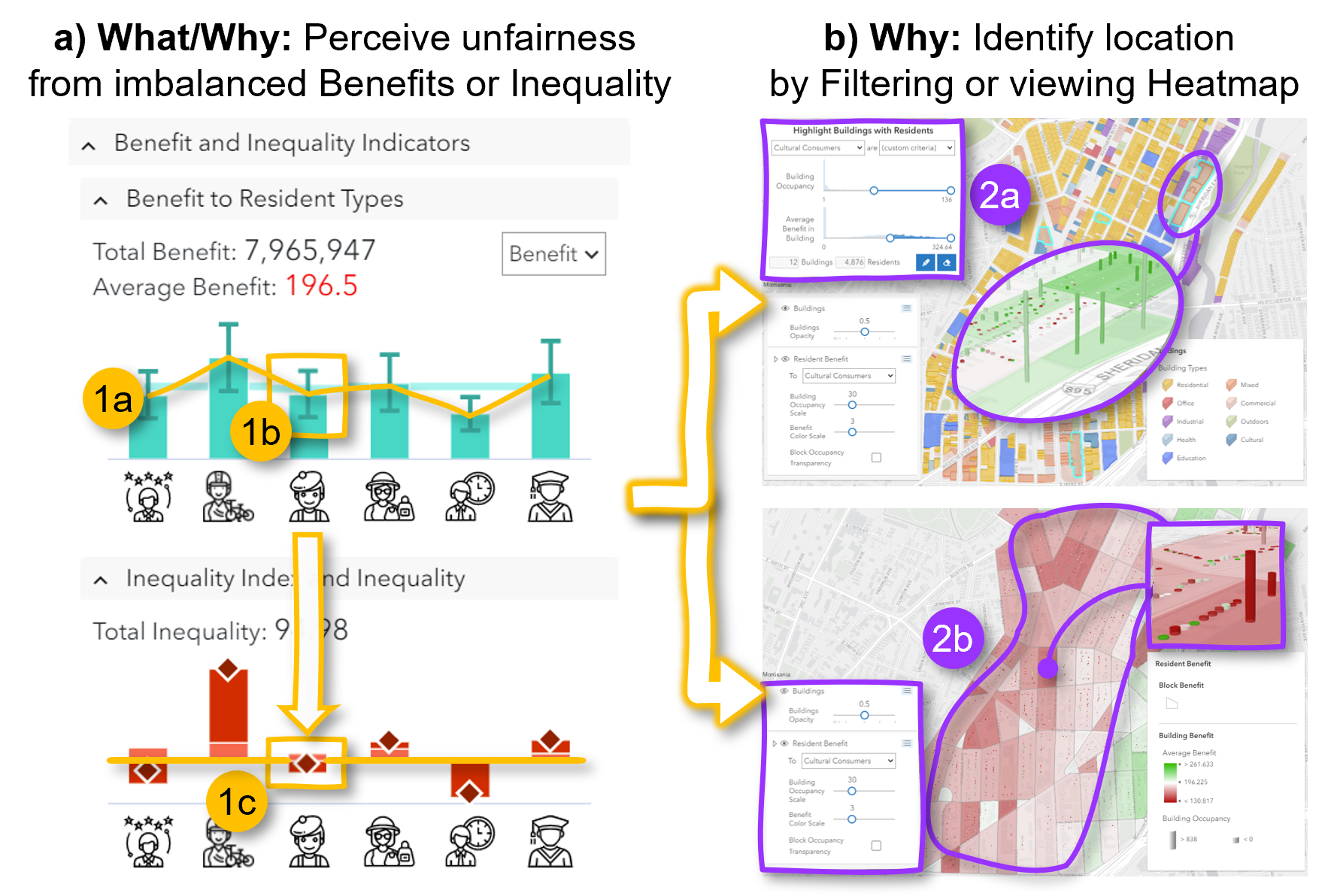}
\captionof{figure}{
User flow to evaluate fairness starts with perceiving unfairness by noting uneven benefits across resident types (Step 1a), more deeply examining one group (1b), and seeing how big the inequality is (1c).
Next, identify the location of inequality by either filtering for buildings with higher occupancy and deviating benefits (2a), or locating dark blocks and buildings in the relative benefits heatmap (2b).
}
\vspace{-0.10in}
\label{fig:CaseStudyPerceiveIdentify}
\end{figure}

\begin{figure}[t]
\centering
\includegraphics[width=0.48\textwidth]{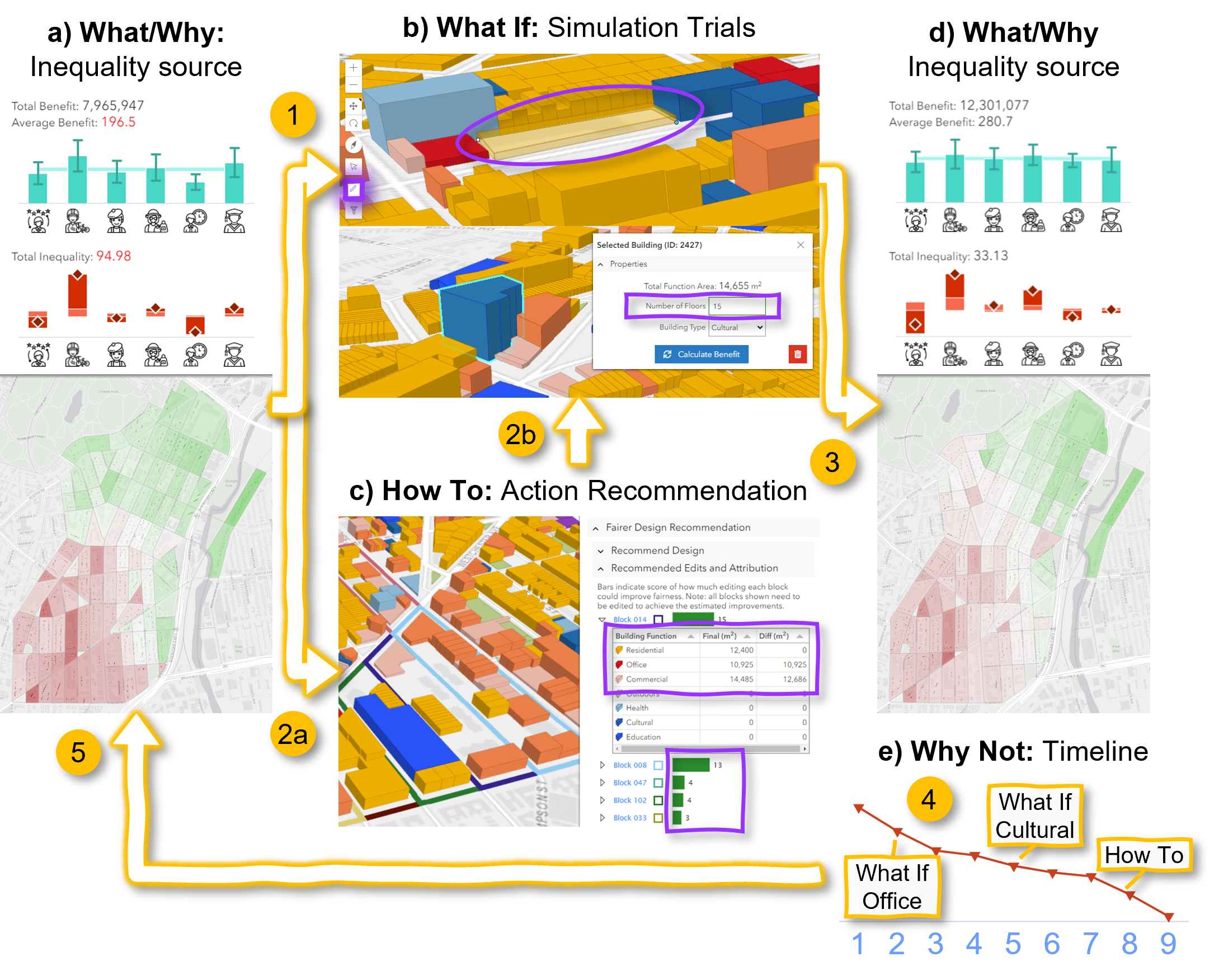}
\captionof{figure}{
User flow to mitigate inequality iteratively with two approaches --- through simulation trials by adding or editing buildings (1), or using the recommendation system (2a) and editing buildings to satisfy the recommendations (2b).
A fairer urban design will have more balanced benefits between and within groups and across locations in the heatmap and smaller group inequality indicators (3).
The user can track improvements in inequality through the design timeline (4), and iteratively improve the design (5).
}
\label{fig:CaseStudyMitigate}
\end{figure}

We demonstrate IF-City with a user flow to design the neighborhood to be fairer.
We articulate steps to 
perceive inequality and identify their sources (Fig.~\ref{fig:CaseStudyPerceiveIdentify}), 
identify accessibility causes (Fig.~\ref{fig:IntelligibleFairnessFeatures}a), and 
mitigate inequality (Fig.~\ref{fig:CaseStudyMitigate}).

\textbf{Perceive inequality problems}. 
The user starts by looking at the Benefit Chart and realizes that the benefits are not evenly distributed among the resident types (Step 1a); specifically, Office Workers, General Consumers and Culture Consumers were relatively disadvantaged, while Outdoor Recreationists had much more benefit than others. 
In this case, she next focuses on the Cultural Consumers (Step 1b) and examines their group inequalities in the Inequality Chart (Step 1c). The between-group (dark red) and within-group (pink) inequalities were particularly small, but she still wishes to examine further. 
Thus, she has perceived some inequality for the Cultural Consumer resident type. The next step is to identify the source of inequality. 

\textbf{Identify causes of inequality}.
There are two methods to identify the locations of greatest inequality --- highlighting filtered buildings or viewing the benefits heatmap.
Using the Building Filtering feature (Step 2a), she selects settings to highlight buildings which have high occupancy ($\ge$50) of Cultural Consumers with high average benefit ($\ge$160 points). These buildings are highlighted in cyan in the 3D map. The user can also review her selections with the Benefits Heatmap and see that these buildings have tall green cylinders.
Alternative to filtering, using the Benefits Heatmap (Step 2b), the user can perceive the uneven distribution of benefits across locations, and focus on the blocks with lowest benefit (darkest red) and high occupancy.

Having identified residential buildings with very unequal benefits, she next finds opportunities to add or remove Cultural spaces.
For locations with above-average benefits for Cultural Consumers, she may view their accessibility circle (Fig.~\ref{fig:IntelligibleFairnessFeatures}a), locate Cultural buildings and decide whether to delete them or edit their function.
Conversely, for locations with below-average benefits, she may change buildings to be used for Cultural functions, or add a new one.

\textbf{Mitigate inequality}.
There are two approaches to mitigate inequality --- trial-and-error with What If simulations or following How To recommendations.
We describe a use case of editing different building function types to improve average benefit from 196.5 and decrease total inequality from 94.98 (Fig.~\ref{fig:CaseStudyMitigate}).
Having found an empty land plot in the south, the user draws a new Mixed building (Step 1) with Office and Commercial floor areas to help disadvantaged Office Workers and General Consumers living nearby. After recalculating benefits, the user can perceive changes to the benefits and inequality in the respective charts and heatmap (Step 3). In the Timeline View (Step 4), the user can track a decrease in the Inequality Indicator score (iteration 1 to 2). She continues to make other edits (Step 5). However, after a few iterations, the decrease plateaus (iteration 4).
Next, the user edits Cultural buildings to help Cultural Consumers, for example, increasing the height of one to 15 floors. This decreases Inequality somewhat, and is repeatable for two more iterations until plateauing again (iteration 5 to 7).

At this stage, the user changes strategy to use How To recommendations (Step 2a). She requests a recommendation with constraints to not increase benefits to Outdoor Recreationalists any higher, not decrease benefits to General Consumers, Cultural Consumers, and Office Workers, and limit floor area changes to 10\%. She is recommended to edit 9 blocks (5 shown in Fig.~\ref{fig:CaseStudyMitigate}c), ranked in decreasing attribution order to improve fairness. For Block 014, if she adds 10,925m$^2$ and 14,485m$^2$ of Office and Commercial floor areas, respectively, then she could decrease the Inequality indicator by 15.
Following this recommendation, she draws a new Mixed building at an empty space and achieved a significant decrease in Inequality (Fig.~\ref{fig:CaseStudyMitigate}e, iteration 8). She runs the recommendation one more time and finishes.
Ultimately, she manages to raise the average benefit to 280.7 and reduce total inequality to 33.13.

%% file: Evaluation.tex

\section{Evaluating with domain experts}

We conducted a qualitative study with six domain experts separately to 1)~confirm the scope and priority of fairness in urban planning objectives, and 2) evaluate the usefulness of IF-City in helping experts to understand the cause of inequality problems and mitigate inequality problems.
Three experts are urban planners (E1 and E2 have experience in China, E3 has experience in Europe) and three are urban designers (E4, E5 and E6 have experience in the United States). 
All experts had 3-10 years of experience.

Lasting two hours, the study procedure included: i) a short interview where the expert was asked about his own understanding of and experience with fairness in urban planning, ii) an introduction to the target neighborhood Southern Boulevard, and its planning goals, followed by a usage tutorial of IF-City and a tutorial of our fairness definition, and iii) two connected tasks to identify the causes of inequality and mitigate the inequality. 
With consent, we recorded audio and screen captured interactions with IF-City. We analyzed the recordings in terms of using intelligibility within the three stages of the IF-Alloc framework.
We present our findings on the experts' understanding on fairness in urban planning, their strategies to complete the tasks, and the reported usefulness and usability of IF-City.

\subsection{Priority and scope of fairness in urban planning}
All experts believed fairness is important. Some (E1, E2, E3 and E6) have previously integrated fairness in their designs qualitatively, while others (E4 and E5) have quantified fairness in terms of accessibility and walkability to evaluate existing urban designs. They all agreed that measuring and visualizing fairness in IF-City made it ``\textit{easy to compare the fairness of planning outcomes}'' [E4], and it was ``\textit{nice to see the intermediate effect of fairness after every single change of the design}''[E1]. 
However, they also argued that fairness is not the first planning priority; there were ``\textit{many other important criteria, such as environmental effect and economic effect}'' [E1]. 

Other than equality, the experts also considered equity as an important aspect of fairness. For example, 
E6 mentioned that ``\textit{designs should guarantee the benefit of low-income population, and also the elderly and disabled people}''. 
Regarding grouping residents into types, all experts would typically segment the population by demographics rather than activity preference, but E3 and E5 felt that segmenting by the latter was useful to inform ``\textit{what are their [population group] needs}'' [E3] and to examine the needs of non-traditional groups such as outdoor recreationalists and educators.

\subsection{Perceiving and identifying the inequality sources}\label{sec:identifyCause}

Overall, experts shared the following common steps to find the sources of the inequality: 
1) identify one disadvantaged resident type at a time from the Benefit and Inequality Charts, 
2) study its residents' building function preferences from the Preference Chart, 
3) locate where they live by filtering for buildings with such residents (Fig. \ref{fig:IntelligibleFairnessFeatures}b, Left) and identifying buildings with high occupancy (i.e., tall cylinders in the benefit heatmaps; see Fig. \ref{fig:IntelligibleFairnessFeatures}b, Right), 
4) check accessibility by selecting the building or block and examining the function type floor area distribution of nearby buildings, 
5) edit building properties (e.g., height, function type) to explore whether the changes could decrease inequality indicator. 
Next, we report how the visual components in IF-City helped the experts to perform these steps.

\textbf{Interpreting Benefit and Inequality Charts (Why)}. 
Most experts preferred to use the Benefit Chart to identify which resident type is disadvantaged, as ``\textit{it is simple and very easy to understand}'' [E1]. However, due to positivity bias, they tended to overlook that advantaged resident types contributed to inequality too.
Though somewhat less intuitive than the Benefit Chart, experts could learn more insights from the Inequality Chart.
E5 found that in the the Benefit Chart both Outdoor Recreationalists and Offices Workers \textit{``deviate from the average with the same amount"} above and below mean, respectively; but in the Inequality Chart the more populous \textit{``Recreationalists have the largest between-group inequality score, so they contribute most to the total inequality"}.
Moreover, while the experts could perceive within-group inequality, they did not investigate their cause further.

\textbf{Locating disadvantaged resident types (Why)}.
Most experts preferred to use the Benefits Heatmap to locate disadvantaged resident types on the map, as they could ``\textit{quickly locate the blocks in red color and with high cylinders}'' [E1]. 
In contrast, Building Filtering ``\textit{[took] more time}'' [E2] and required more interactions to tune settings for a suitable selection threshold. 
Nevertheless, it provided detailed distributions of benefit and occupancy, allowing experts to select the buildings based on perceived quantiles.

\textbf{Identifying causes of inequality due to accessibility (Why)}. On finding where disadvantaged residents were on the map, experts further examined the accessibility at the affected buildings. Most experts only quickly viewed the blue Accessibility Circle to identify what function types it encloses; only a few experts studied the Accessibility Chart to examine the floor area distribution of function types. This helped them to clearly see how much specific function types could be added or removed, e.g., E1 pointed out ``\textit{the bar chart helps me to compare floor area of different function types and find out which function type should be added}``. 

\subsection{Mitigating and verifying inequality}

Here, we describe how experts used two approaches to mitigate inequality, and verified improvements while iterating.

\textbf{Trial and error strategy (What If)}. 
All experts first chose to manually improve fairness by editing buildings or blocks identified as disadvantaged, and running the What If simulation to recalculate the benefit and inequality indicators.
Most experts added new Office, Culture and Mixed buildings to improve the benefit for Office Workers, Culture Consumers and General Consumers. 
They tended to add new buildings than edit existing ones to avoid disrupting existing usages and activities.
The edits were interspersed with studying intelligibility features in the Benefit and Inequality Charts across resident types, the residents' function type preferences, and where they lived by seeing the Benefit Heatmap. This helped E2 and E4 to retarget their edits.
With their independent effort, four experts (E1, E4, E5, E6) achieved the planning goal of sufficiently low inequality and high total benefit, but they took a long time to do so (40-90 minutes).
To further improve fairness, the experts subsequently used the How To recommendations.

Finally, enunciating the wicked nature of urban planning, the experts reported that mitigating inequality was complex and challenging because improving benefits for one resident type may hurt others. 
It would be \textit{``tedious to check which building I can change so that no one's benefit gets hurt''} [E2]. 
E6 found that the benefit of General Consumers ``\textit{is hard to improve because their preference is multi-fold}'', to include Office, Commercial and Cultural buildings, but changing them will also affect the benefits to Office Workers, Commercial Consumers and  Culture Consumers.

\textbf{How To recommendations}.
After manual attempts to improve fairness, the experts investigated further improvements with automatic recommendations.
Most experts were conservative and wanted to see the recommendation details to examine whether they were realistic and feasible. 
E2, E3, E5 and E6 were interested in how the recommended edits were automatically calculated. 
Conversely, E1 and E4 were initially skeptical and assumed that the calculations were too simple; they believed that planning is complicated and many factors should be considered such as street views and height control.
After following and executing the top few recommendations, all experts found the recommended edits were feasible and led to large decreases in the inequality indicator. 
They ``\textit{trust[ed] the recommendations}'' [E5] and found that the recommendation table was ``\textit{very transparent}'' [E3] as they learned ``\textit{which block needs what [to edit]}'' [E2]. The recommendations helped them to ``\textit{narrow down the inequality problems to certain blocks}'' [E4]. 
They also appreciated coarse the block-level recommendation as they ``\textit{[had] the freedom to make detailed design within a block}'' [E6]. E3 was an exception who wanted the recommendation ``\textit{automated into buildings}'' so that he did not need to make any edit by himself.
Regarding constraining the recommendations, the experts appreciated the control over prioritizing resident types, as ``\textit{sometimes urban planners need to make sure the benefit of certain resident types should be improved by some policy}'' [E3]. They also liked the setting to threshold the floor area change because they could easily limit a planning budget. 
Overall, the experts thought the How To recommendation feature was ``\textit{a good guide}'' [E3] and wanted to use it rather than solving the inequality problems by themselves manually.

\textbf{Verifying and inspecting improvements (What (comparison) \& Why Not)}. 
Most experts used the Timeline View to check whether the inequality indicator improved with each design iteration. 
E1 liked that he could perceive \textit{"the intermediate effect of fairness after every single change of the design"}. E2 appreciated perceiving the \textit{"magnitude and scale"} of his edits on the inequality index. This indicates the importance of the comparative use of What, to contextualize the meaning of the inequality index that would otherwise be too abstract.
Furthermore, 
E1 clicked on her past saved designs and compared details with the current version. This helped her to understand that her adding of Office buildings in the south and Commercial buildings in the middle of the neighborhood had significantly improved fairness.

\subsection{Feedback on usability and usefulness}

\textbf{Usability}. All the experts found the Benefit and Inequality charts,  the Heatmap, and Cylinders easy to understand:  ``\textit{These features are very helpful to see the inequality situation. I like this visualization and I think it is very presentable to clients and stakeholders}'' [E4]. E6 liked the design of IF-City and felt it easy to get familiar with it as ``\textit{this tool is quite intuitive}''. E5 commented that the ``\textit{information is well organized, and the concept of fairness is clearly conveyed}'', and believed planners could learn it in a short time.
However, she would prefer less scrolling to see all the charts and tables.

\textbf{Usefulness}. The experts reported IF-City would be useful for ``\textit{overall high-level land use planning}'' [E4], ``\textit{before the block subdivision process}'' [E6] and for ``\textit{[re-evaluating] land use}'' [E4]. E4 also commented that having the quantitative tool is helpful to explicitly articulate fairness as ``\textit{most of the time the fairness was discussed in conversation with photos}''. E5 thought IF-City could be integrated with the existing professional design tools such as Autodesk and Rhino.

%% file: Discussion.tex
\section{Discussion}

We have demonstrated the usability and usefulness of intelligibility for fair urban design. Here, we discuss 1)~the need for domain-specific fairness visualizations, 2)~the usefulness of intelligibility in fairness understanding, and 3)~using IF-City to analyze fairness for different neighborhoods or cities, 4)~generalizing for different notions of fairness, and 
5)~generalizing IF-Alloc framework beyond urban planning.

\textbf{Domain-specific fairness visualizations.}
Unlike data scientists whose role is to understand and model data, expert users have domain-specific roles and tasks with specialized tools and workflows. 
Towards this end, we have identified fair allocation as a concern for urban planners, defined and implemented inequality indicators and explanations for resident types and locations, designed visualizations of group fairness, and integrated fairness visualizations seamlessly into a workflow for urban design. 
Our evaluation users were comfortable using the tool, appreciated the need for fairness, and were effective to reduce inequality.
Though important, fairness is not the primary goal in many applications, so its visualization needs to be carefully designed and integrated.
We drew from existing urban planning tools to leverage geospatial visualizations, 3D models, and accessibility calculations to develop a domain-specific intelligible fairness tool.
Other applications that can benefit from further study include 
social network analysis~\cite{henry2007nodetrix} or computer networking~\cite{goodall2005user} by visualizing graph networks, 
and fair scheduling~\cite{levin2010dp} by visualizing time tables.

\textbf{Intelligibility for fair allocation.}
We have developed various intelligibility features to support three stages to iteratively perceive, understand, and mitigate inequality in resource allocation.
``Why" explanations to identify the cause of inequality were useful, especially with heatmaps to locate inequality, perhaps due to the familiarity with maps in urban planning.
Providing both What If simulation and How To recommendation to mitigate inequality helps to empower expert designers to make fine-grained decisions, while supporting efficient suggestions on-demand. This grants control to domain experts to integrate other concerns or constraints in their solutions.
In our study, experts first explored via trial-and-error (What If), before exploring automatic recommendations.
However, for cases where users are not expert planners (e.g., \cite{lee2019webuildai}), it may be better to prioritize How To recommendations.
Finally, although contrastive explanations are key for explainability goals~\cite{miller2019explanation}, we found limited use of the Why Not feature. Our experts did use the What (Comparison) timeline to track their progress, but rarely compared details between past and current designs. Perhaps, due to them easily remembering their earlier designs, the tediousness of examining many buildings between urban plans, and the brief duration of the experiment session. We expect Why Not to be more useful in usages spanning hours or days.

\textbf{Generalizing to analyze fairness for other neighborhoods or cities}. 
Using IF-City, urban planners can upload the designs of other cities (as Geo-JSON) and their resident demographic or preference data (JSON) to assess the fairness of those cities. 
Moreover, by loading the dashboard in multiple web browser windows, users can examine detailed reasons why one city may be fairer than another. 
We have evaluated IF-City on a small neighborhood and note the need to scale the calculations and memory requirements for much larger cities. 
Furthermore, while we focused on fairness, 
IF-City can be extended to other objectives, such as green space and economic sustainability.

\textbf{Incorporating both equity and equality for fair cities}.
IF-City balances resident benefits to improve the \textit{equality} of outcome improvements across groups.
However, some groups may be tacitly disadvantaged prior to allocation, and it may be fairer to give them further aid to achieve \textit{equity} in benefits instead. 
Our evaluation experts mentioned that the benefit
to the disadvantaged (e.g., elderly, children, low-income) should take priority over others.
In IF-City, we can define a \textit{population priority weight} $\rho_g$ for each different resident type $g$ to prioritize or penalize their benefits, i.e., $b_i \leftarrow \rho_g b_i$. The weighted benefits can then be substituted into the calculations for inequality, and resident allocation.

\textbf{Generalizing to other fairness applications beyond urban planning.}
IF-Alloc can be applied to other fair allocation applications, such as job allocation for the gig economy~\cite{fieseler2019unfairness}, product placement in retail or online stores~\cite{yoo2009effects}, worker shift allocation (e.g., in restaurants, hospitals)~\cite{uhde2020fairness}, and donation division~\cite{lee2019webuildai}.
Following the steps in Fig.~\ref{fig:SystemOverview}, we describe an example application for fair driver job allocation in ride-sharing applications~\cite{suhr2019two, xu2020trading}.
1) Extract map and road network data to determine routes and distances.
2) Extract driver background and behavioral information of drivers from worker surveys, app usage, and driving trajectories (2a) to cluster (2b) them into driver types (2c) based on their demographics, preferred driving times and locations, etc. (2d).
3) Simulate a job request scenario (multiple riders' requests, given multiple drivers) and automatically allocate jobs~\cite{xu2018large}.
4) Calculate the benefits (earnings) for each driver given their job allocation, aggregate benefits across driver types (4a) and current locations (4b), and calculate group inequalities using Generalized Entropy (4c).
Steps 4a and 4b support explaining Why about the fairness.
5) Enable job planners to adjust planning parameters (e.g., commission fee~\cite{raman2021data}, surge pricing rate~\cite{pandey2021disparate}, incentive threshold~\cite{lu2020say},
working speed~\cite{lee2015working})
and run the allocation engine to simulate What If the parameters were different (5a), or request recommendations of fairer settings (5b).
In general, IF-Alloc can be applied to applications involving resource or task allocation, planning parameters, and diverse stakeholders.

%% file: Conclusion.tex
\section{Conclusion}

We have proposed the Intelligible Fair Allocation (IF-Alloc) Framework to aid 
domain experts to examine fairness in allocation in domain-specific tasks. 
The framework supports different intelligibility features 
for three stages to perceive fairness, identify inequality causes, and mitigate inequality.
We applied the framework towards fair urban planning 
to develop the Intelligible Fair City Planner (IF-City) interactive visual tool.
IF-City provides Why explanations to trace and attribute inequality to resident types and locations, What If explanations to simulate the impact of new designs on fairness, How To explanations to automatically generate fairer recommendations, and Why Not explanations to compare design details across iterations. 
We demonstrated and evaluated the usage and usefulness of IF-City on a real neighborhood in New York City, US, with practicing urban planners and urban designers. 
Using various intelligibility features, urban planners could perceive and identify causes of inequality, and mitigate inequality. 
This works sheds light on how to carefully design detailed user interaction features for fair design in collaborative human-AI planning.